\documentclass{article}
\usepackage{arxiv}
\usepackage{cite}
\usepackage{amsmath,amssymb,amsfonts}
\usepackage{graphicx}
\usepackage{hyperref}
\usepackage{caption}
\usepackage{subcaption}
\usepackage{textcomp}
\def\BibTeX{{\rm B\kern-.05em{\sc i\kern-.025em b}\kern-.08em
    T\kern-.1667em\lower.7ex\hbox{E}\kern-.125emX}}
\usepackage{scalerel}
\usepackage{cleveref}
\usepackage{multirow}
\usepackage{makecell}

\usepackage{algorithm}
\usepackage{algpseudocode}
\usepackage[algo2e]{algorithm2e}
\usepackage{xcolor}

\definecolor{gfet}{HTML}{dcd62f}
\definecolor{gfem}{HTML}{80DC2F}

\newcommand{\argmin}[2]{\operatornamewithlimits{argmin}\limits_{#1}{\;\;#2}}

\def\*#1{\mathbf{#1}}   

\def\L{{\cal L}}

\newcommand{\genm}[3]{
	\left[\begin{array}{cc}
		#1  &  #3 \\
		#3 & #2 
	\end{array}
	\right]
}

\newtheorem{prop}{Proposition}

\usepackage{enumitem}

\usepackage{mwe} 
\usepackage{xcolor}

\def\L{{\cal L}}

\def \r{\mathbf{r}}

\def \ah{\hat{\mathbf{a}}}
\def \zh{\hat{\mathbf{z}}}
\def \a{{\mathbf{a}}}
\def \z{{\mathbf{z}}}

\def \K{\mathbf{K}}
\def \L{\mathbf{L}}

\definecolor{gfet}{HTML}{dcd62f}
\definecolor{gfem}{HTML}{80DC2F}

\title{Generalized Hessian-Schatten Norm Regularization for Image Reconstruction}

\author{  
		Manu Ghulyani$^*$, \;\; Deepak G Skariah\thanks{Manu Ghulyani and Deepak G Skariah have equally contributed to the work},   \;\; 
Muthuvel Arigovindan\thanks{Corresponding author (mvel@iisc.ac.in)} \\
		Imaging Systems Lab\\  
		Department of Electrical Engineering\\
		Indian Institute of Science (IISc)\\ Bangalore 560012, India\\
				}

\begin{document}
	
%
%
%

\maketitle

\begin{abstract}Regularization plays a crucial role in reliably utilizing imaging systems for scientific and
	medical investigations. It helps to stabilize the process of computationally undoing  any degradation caused
	by  physical limitations of the imaging process. In  the past decades,  total variation regularization,
	especially second-order total variation (TV-2) regularization played a dominant role in the literature.
	Two forms of generalizations, namely Hessian-Schatten norm (HSN) regularization, and total generalized
	variation (TGV) regularization, have been recently proposed and have become significant developments in
	the area of regularization for imaging inverse problems owing to their performance. Here, we develop
	a novel regularization for image recovery that combines the strengths of these well-known forms. We
	achieve this by restricting the maximization space in the dual form of HSN in the same way that TGV
	is obtained from TV-2. We name the new regularization as the generalized Hessian-Schatten norm
	regularization (GHSN), and we develop a novel optimization method for image reconstruction using the
	new form of regularization based on the well-known framework called alternating direction method
	of multipliers (ADMM).
	We demonstrate the strength of the GHSN using some reconstruction examples.
\end{abstract}

Regularization, Total Variation, Total Generalized Variation, Hessian-Schatten Norm, Inverse Problems,  MRI Reconstruction.

\section{Introduction}
\label{sec:introduction}


Images acquired using different imaging devices are  inevitably corrupted 
due to the physical limitations of the image formation model. 
An estimation 
scheme \cite{papoulis2002probability} that employs knowledge of 
the image formation forward model to generate a better quality 
estimate is known as image restoration/reconstruction \cite{banham1997digital}. 
The relative improvement in the image quality  obtained by image
restoration/reconstruction is typically significant in most modalities 
in general,  and in particular, in modalities such as
MRI imaging 
\cite{deriche2003half,ramani2012regularization}, computed tomography 
\cite{rathee1992image, ma2011low}, confocal microscopy 
\cite{dey2004deconvolution}  and widefield microscopy 
\cite{arigovindan2013high}.
One of the classical approaches to image restoration is the regularized
approach \cite{hansen2010discrete}. It formulates the required reconstruction
by means of the solution to the optimization problem as given below:
\begin{equation}
{g}_{opt} = \argmin{g}{J(g) = \big[ G(m,g) + \lambda R(g)\big ]},
\end{equation}
where  $G(m,g)$   is the data fitting functional that measures 
the goodness of fit of the candidate image $g$,  to the measured image $m$,  and 
$R(g)$  is the regularization functional that measures 
some kind of roughness of the image. The structure of data fitting
term is  dependent on the forward model of the imaging device and
the assumed statistical model of the noise.  The regularization 
functional \cite{kang1995general} represents the prior information
we have about the class of images we are trying to restore. The
constant term $\lambda$ is a user  parameter that allows
a trade off between the  regularization term and 
the data fitting term.

Design of  regularization functional   $R(g)$ is a long-studied
problem in signal and image processing \cite{banham1997digital}. One of the classic
regularization techniques is Tikhonov regularization \cite{tikhonov1977solutions}. 
Tikhonov regularization was originally employed for solving integral 
equations, but it proved to be successful in imaging inverse problems 
\cite{ying2004tikhonov} as well. The earliest form of  Tikhonov regularization 
was constructed using  sum-of-squares of the candidate image/signal \cite{tikhonov1977solutions},  and
then the later forms were constructed using sum-of-squares of image/signal derivatives.
Derivative based Tikhonov regularized reconstruction can be
expressed as
\begin{equation}
\begin{gathered}
\label{eq:tikh1}
g_{opt} = \underset{g}{\operatorname{argmin}}\;\;G(m, g) + \lambda
\underbrace{\sum_{\mathbf r} \|({\mathbf d}*g)({\mathbf r}) \|_2^2}_{R_f(s)}  \\
\mbox{with}\;\; d(\r) = [d_{x}(\r), d_{y}(\r)]^{T},
\end{gathered}
\end{equation}
where $d_{x}(\r)$ and $d_{y}(\r)$ are filters implementing first order derivatives,
$\frac{\partial}{\partial x}$, and $\frac{\partial}{\partial y}$ respectively.
The main advantage of using Tikhonov regularization is that the required reconstruction
can be expressed in terms of linear system of equations when the data fitting term
is also quadratic.  However, Tikhonov regularization
leads to over-smooth solutions. Nevertheless,  Tikhonov regularization
is still widely used epspecially in large scale problems such 3D deconvolution
\cite{tikh2}. 
It was generally believed that non-linear methods gave better quality reconstructions,
and the earliest forms of regularizations that led to non-linear methods
were based on $l_1$ norm of the wavelet transform of the candidate image.  Representative
methods in this category include 
ISTA \cite{chambolle1998nonlinear,
	figueiredo2003algorithm,daubechies2004iterative}, FISTA \cite{beck2009fast} 
and TwIST \cite{nTWIST}.  These methods are 
primarily based on the fact that wavelet transform of typical images have low
$l_1$ norm.  
The basis used in the wavelet transform can be adapted to data to give better performance in 
image restoration problems \cite{bresler_dict}.  It should be emphasized that
the factor that makes wavelet regularization better than the Tikhonov is that
the wavelet transform has inbuilt derivatives and hence wavelet regularization
in some sense becomes equivalent to $l_1$ minimization image derivatives. This leads
to better preservation of structures than Tikhonov regularization
which smooths out edges due to $l_2$ minimization of derivatives.

The seminal work by 
Rudin, Osher and Fatemi demonstrated that  minimization of norm of image derivatives,
in particular gradients, leads to better preservation of image structures  \cite{rudintv}.
This regularization is called 
total variation (TV) regularization and its 
practical 
and theoretical advantages have been well demonstrated 
\cite{fessler2019optimization, poon2015role,needell2013stable}. This 
makes total variation an active area of research even after decades of 
discovery. TV regularized image reconstruction 
can be expressed as
\begin{equation}
\begin{gathered}
\label{eq:tv1}
g_{opt} = \underset{g}{\operatorname{argmin}}\;\;G(m, g) + \lambda
\sum_{\mathbf r} \|({\mathbf d}*g)({\mathbf r}) \|_2 .
\end{gathered}
\end{equation}
For a rigorous treatment of Total Variation (TV), the reader
can refer to \cite{chambolle2010introduction}.  
TV regularization  has been widely used \cite{TV_App_BlindDecon, TV_App_Microwave,
	TV_App_Seismic, TV_App_Wavelet, TV_App_Wlinpaint}
because of  its ability to recover sharp image features in the
presence of noise and in the cases of under-sampling. 
While TV can retain edges \cite{rudintv} in the reconstruction as compared to 
Tikhonov regularization \cite{chambolle2010introduction}, and in particular,  it can recover sharp jumps in the
reconstruction even in the presence large amount of noise and/or undersampling.
At the same time it has a disadvantage that, in the presence of large amount of noise and/or undersampling,
it approximates smooth intensity variations in terms of piece-wise constant segments, which is known 
as   staircase artifacts \cite{chan2000high}.   
Higher order extensions of TV  \cite{chan2000high} have been proposed to avoid staircase effect and they
deliver better restoration.
Second order TV (TV-2)  \cite{dogan2011second,Scherzer_tv2_98,steidl2006note} restoration was proposed as 
\begin{equation}
\begin{gathered}
\label{eq:tv2}
g_{opt} = \underset{g}{\operatorname{argmin}}\;\;G(m,g) + \lambda
\underbrace{\sum_{\mathbf r}\|({\mathbf d}_s*g)({\mathbf r})\|_2}_{R_s(g)}, \\
\mbox{with}\;\; {\mathbf d}_s({\mathbf r}) = [d_{xx}({\mathbf r})\;\;d_{yy}({\mathbf r})\;\;\sqrt{2}d_{xy}({\mathbf r})]^t,
\end{gathered}
\end{equation}
where $d_{xx}(\r), d_{yy}(\r)$, and $d_{xy}(\r)$ are discrete filters implementing second order derivatives 
$\frac{\partial^2}{\partial x^2}$, $\frac{\partial^2}{\partial y^2}$ and $\frac{\partial^2}{\partial x\partial y}$ 
respectively. TV-2  recovers linear intensity variations in the presence large amount of noise and/or 
undersampling;  it looses the ability to reproduce sharp intensity jumps that TV-1 can recover.
Another second-order derivative based formulation is Hessian-Schatten (HS) norm regularization  \cite{hessian},
which has been proposed as a generalization
of the standard TV-2 regularization.  It is constructed as an $\ell_p$ norm of eigenvalues of the Hessian
matrix, which
becomes the standard TV-2 for $p=2$.  HS norm with $p=1$ has been proven to yield best resolution
in the reconstruction, since this better preserves eigenvalues   of the Hessian  \cite{hessian}.
Let 
${\mathbf H}(\r)=
\genm{d_{xx}(\r)}{d_{yy}(\r)}{d_{xy}(\r)}$
and let ${\cal E}(\cdot)$ be the operator that returns the vector containing
the eigenvalues of its matrix argument. Then HS norm regularization of order $p$ is expressed as
\begin{equation}
\label{eq:hs_reg}
g_{opt} = \underset{g}{\operatorname{argmin}}\;\;G(m, g) + \lambda
\sum_{\mathbf r}\left\|{\cal E}(({\mathbf H}*g)({\mathbf r}))\right\|_p.
\end{equation}
Since the eigenvalues are actually directional second derivatives taken along principle
directions, setting $p=1$ better preserves the local image structure.

 Papafitsoros  et al. proposed a regularization method that combines
 both first- and second-order derivatives \cite{Combined_order_TV}.
The reconstruction problem is formulated as given below,
\begin{equation}
g_{opt} = \underset{g}{\operatorname{argmin}}\;\;G(m,g)
+ \lambda \alpha_f R_f(g) + \lambda \alpha_s R_s(g),
\end{equation} 
where $\alpha_f$  and $\alpha_s$  are user parameters that 
determine the relative weight.
 A generalization for total variation to higher order terms, named as 
total generalized variation (TGV)  has also been proposed  \cite{tgv}.  It is 
generalized in two  ways:  it is formulated for any
general derivative order;   for any given order,  it is generalized in the way
how the derivatives are penalized.  The second form of generalization is obtained
by expressing the standard total variation regularization in dual form  as 
a maximization, and then by imposing spatial smoothing constraint in the maximization problem.
The second aspect of generalization  has  more significant impact in practical
point of view.  In fact,  the version of  TGV regularization that has been 
applied for non-trivial inverse problems (problems other than denoising) 
obtained by restricting the maximum order to be two; compared to most widely
used TV-2 regularization,  it differs only by the second aspect of generalization.
This form of regularization is called second order TGV (TGV-2)  and takes
the following form:
\begin{equation}
\label{eq:tgv2imp}
TGV_2(g, \alpha_f, \alpha_s)   = \underset{{\mathbf u}}{\operatorname{argmin}}\;\;
\alpha_f\sum_{\mathbf r} \|({\mathbf d}*g)({\mathbf r})-{\mathbf u}({\mathbf r})\|_2 
+  
\frac{1}{2} \alpha_s \sum_{\mathbf r}  \|({\mathbf d}*{\mathbf u}^t)({\mathbf r})+({\mathbf u}
*{\mathbf d}^t)({\mathbf r})\|_F,
\end{equation}
where ${\mathbf u}({\mathbf r})$ is an auxiliary $2\times 1$ vector image.
The TGV-2 functional is able to spatially adapt to the underlying image structure 
because of the minimization w.r.t.
auxiliary variable
${\mathbf u}$. Near edges, ${\mathbf u}({\mathbf r})$ approaches zero leading to TV-1-like 
behaviour which allows sharp jumps in
the edges.  On the other hand,  in smooth regions,  ${\mathbf u}({\mathbf r})$ approaches  
${\mathbf d}*g({\mathbf r})$ leading
to TV-2-like behaviour which will avoid staircase artifacts. This means that TGV-2 
combines the best of TV-1 and
TV-2 regularizations.  However, the drawback with TGV
functional is that   the weights  $\alpha_f$  and $\alpha_s$ have to be chosen 
by the user.



The recent leap in the computing  power of desktop computers led  the application
deep neural networks (DNN) for  image restoration 
\cite{ker2017deep}.  These DNN based image restoration methods can be categorized 
into two types. In the first category, a map (composed of multi-layer neural networks)
that can estimate underlying image from  measured image 
is learned   from the several training pairs, where each pair has a measured image
and the underlying image that generated the measured image.
Note that this map plays dual role:  it  accounts for stable inversion of
 imaging forward model,  and encompasses prior knowledge of typical
 ground truth images. This type of DNN methods were 
applied for MRI image restoration \cite{dagan,deepmri}, CT image \cite{fbp} 
reconstruction, image deconvolution \cite{xu2014deep} and other imaging inverse 
problems. In the second category,   only the effect of noise is 
learned. Therefore, these maps are known as denoisers and plugged as a module in a 
variable splitting optimization scheme (e.g., Primal-Dual splitting).  Some prominent 
works of this category are \cite{pnp1} and \cite{pnp2}. The main argument that supports
the use of DNN is that, while regularization methods impose adhoc prior beliefs on
the image to be restored,   a trained DNN encompasses a more natural knowledge
resulting from training data. However, 
a  recent work by Hansen et al. \cite{antun2020instabilities} shows that 
such learned maps can be unstable. Further,  need for large amount of training
data limits the applicability of DNN based methods.

In this paper,  we develop a novel type of regularization by generalizing the 
Hessian-Schatten norm regularization in the same way that TGV-2 regularization
generalizes the TV-2 regularization.  The resulting form of  regularization includes
TV-2,  TGV-2, and Hessian-Schatten norm regularization as special cases.
We call this regularization as the generalized Hessian-Schatten norm regularization
(GHSN).  Next,  we develop a novel optimization method for image reconstruction
using the GHSN  and demonstrate the effectiveness of  GHSN regularization   using  
numerical experiments involving  reconstructions  from sparse Fourier samples.

\section*{Notations and mathematical preliminaries}{\label{notation}}

\begin{enumerate}
	
	\item
	Vectors are represented by lower-case bold faced letters with the elements
	represented by the same letter with a subscripted index.  For example,
	${\mathbf v}$ denotes a vector and its $i$th element is denoted by $v_i$.
	For a vector
	${\mathbf v}$,   $\|{\mathbf v}\|_p$ denotes the
	$l_p$ norm given by  	$\|{\mathbf v}\|_p=(\sum_{i=1}^n |\mathbf v_i|^p)^{1/p}$. It can be shown
	that $\|{\mathbf v}\|_p$ with $p\rightarrow \infty$ converges to the component
	of ${\mathbf v}$ that has the largest magnitude, and hence we write
	$\|{\mathbf v}\|_\infty = \max_i \{{|\mathbf{v}_i|:i=1,2,...,n}\}$.
	\item
	We will deal
	with vector images;  vector images are discrete 2D arrays where each pixel
	location has a vector quantity. It is denoted by lower-case bold-faced letter
	with a bold-faced lower-case letter as an argument.  For example,
	${\mathbf v}({\mathbf r})$  is a vector image with ${\mathbf r}=[x\;y]^t$ representing a 2D pixel
	location. Depending on the context, the symbol denoting the pixel location
	may be omitted.
	\item
	For a vector image, 	${\mathbf v}({\mathbf r})$,  $\|{\mathbf v}\|_{1,2}$
	denotes  $\|{\mathbf v}\|_{1,2}=\sum_{\mathbf r}\|{\mathbf v}({\mathbf r})\|_{2}$. It is
	the  sum of  pixel-wise $l_2$ norms, where $\sum_{\mathbf r}$ denotes the sum
	across pixel indices.  Throughout the paper, we do not specify the bounds
	of summation as it is always the same and ranges from the first to last pixels.
	The norm $\|{\mathbf v}\|_{1,2}$ is a composition of two norm, and is called the
	mixed-norm.
	\item
	Matrices are represented by upper case bold faced letters.  
	For a matrix ${\mathbf M}$,  $\|{\mathbf M}\|_{F}$  denotes the Frobenius norm,
	which equal to the square root of sum of squares of its elements; it can
	be written as $\|{\mathbf M}\|_{F}=\sqrt{Tr({\mathbf M}^t{\mathbf M})}$, where $Tr(\cdot)$
	denotes the summing of diagonal elements.
	Next $\|{\mathbf M}\|_{S(p)}$ denotes
	the Schatten p-norm  of the matrix, which is the
	$l_p$ norm of the vector of  singular values  of the matrix ${\mathbf M}$. 
	For a symmetric matrix,  it is also the same as the $l_p$ norm of the Eigen values.
	In other words,
	if ${\cal E}$ denotes the operator that returns the vector of eigenvalues of
	a matrix,  then $\|{\mathbf M}\|_{S(p)} =   \|{\cal E}({\mathbf M})\|_p$.
	\item
	We will also deal
	with matrix images;  matrix images are discrete 2D arrays where each pixel
	location has a matrix quantity. It is denoted by upper-case bold-faced letter
	with a bold-faced lower-case letter as an argument.  For example,
	${\mathbf M}({\mathbf r})$  is a matrix image with ${\mathbf r}$ representing a 2D pixel
	location. Depending on the context, the symbol denoting the pixel location
	may be omitted.  For a vector image,  $\mathbf{v}({\mathbf r})$,  its $j$th scalar image
	is given by $v_j(\mathbf{r})$.   For a matrix image,  $\mathbf{M}({\mathbf r})$,  its $(i,j)$th scalar image
	is given by $m_{i,j}(\mathbf{r})$.   
       \item
	For  an matrix image, ${\mathbf M}(\r)$, let $\|{\mathbf M}\|_{1,S(p)}$  denote
	the $l_1$ norm of the pixel-wise Schatten p-norms.  In other words,  we
	have $\|{\mathbf M}\|_{1,S(p)}=\sum_{\r}\|{\mathbf M}(\r)\|_{S(p)}$. Further, 
	let $\|{\mathbf M}\|_{\infty,S(p)}$  denote
	the $l_\infty$ norm of the pixel-wise Schatten p-norms.  In other words,  we
	have $\|{\mathbf M}\|_{\infty,S(p)}=\max_{\r}\|{\mathbf M}(\r)\|_{S(p)}$. 
	Note that the  norms $\|{\mathbf M}\|_{1,S(p)}$ and $\|{\mathbf M}\|_{\infty,S(p)}$ are
	mixed-norm.
	\item
	For square matrices of same dimension,  ${\mathbf M}$,   ${\mathbf N}$, let  
	$\langle {\mathbf N}, {\mathbf M} \rangle = Tr({\mathbf N}^t{\mathbf M})$.
	For a square matrix ${\mathbf N}$  let $\overline{\mathbf N}=\frac{{\mathbf N}+{\mathbf N}^t}{2}$.
	Then the following holds for any pair of square matrices:  
	$\langle  {\mathbf M}, \overline{\mathbf N}\rangle = \langle  \overline{\mathbf M}, {\mathbf N}\rangle$.
	\item
	If 
	${\mathbf M}$,  and  ${\mathbf N}$  denote  matrix images such that
	for each pixel index ${\mathbf r}$,  ${\mathbf M}({\mathbf r})$,  and  ${\mathbf N}({\mathbf r})$
	denotes square matrices of the same dimensions, 
	then,  $\langle {\mathbf N}({\mathbf r}), {\mathbf M}({\mathbf r}) \rangle$  denotes
	$Tr({\mathbf N}^t({\mathbf r}){\mathbf M}({\mathbf r}))$.  On the other hand, 
	$\langle {\mathbf N}, {\mathbf M} \rangle$  denotes  $\sum_{\r}Tr({\mathbf N}^t({\mathbf r}){\mathbf M}({\mathbf r}))$.
	\item
	For a matrix image ${\mathbf M}({\mathbf r})$, the norm $\|{\mathbf M}\|_{1,S(p)}$ 
	can be expressed as a maximization
	of inner product of the form $\langle \cdot, \cdot \rangle$ \cite{hessian}.  Specifically we
	can write
	\begin{equation}
	\|{\mathbf M}\|_{1,S(p)} = \max_{\|{\mathbf N}\|_{(\infty,S(q))}\le 1}\langle {\mathbf N}, {\mathbf M} \rangle,
	\end{equation}
	where $q$ is the real positive number satisfying $1/p+1/q=1$. The notation
	$\max_{\|{\mathbf N}\|_{\infty,S(q)}\le 1}$  denote maximization within the set of 
	matrix images with the value of mixed norm $\|{\mathbf N}\|_{\infty,S(q)}$ upper-bounded by $1$.
	\item In this paper, * denotes the 2-D convolution operation, i.e $(x*y)(\mathbf r')=\sum_{\mathbf r}x(\mathbf r)y(\mathbf r'-\mathbf r).$ 
	We extend   notion of convolution  to matrix images  by using the rules of matrix multiplication. 
	To be more specific, let  $\mathbf X$ be an image  of $m\times l$ matrices and
	let  $\mathbf Y$  be image of  $l\times n$ matrices. Then, $(\mathbf X*\mathbf Y)_{i,j}(\r)=\sum_{p=1}^{l}(\mathbf x_{i,p}*\mathbf y_{p,j})(\r)$
	 for $i=1,...,m$.  As an example,  let  both $\mathbf{X}$ and $\mathbf{Y}$ be $2\times 2$ matrix images.
	 Then the elements of   $\mathbf{Z}= \mathbf{X*Y}$,  can be written as 	
	\begin{align*}
	&\mathbf{z}_{1,1}(\mathbf{r}) =   
	(\mathbf{x}_{1,1}*\mathbf{y}_{1,1})(\mathbf{r}) + (\mathbf{x}_{1,2}*\mathbf{y}_{2,1})(\mathbf{r}), \;\;\\&
	\mathbf{z}_{1,2}(\mathbf{r}) = (\mathbf{x}_{1,1}*\mathbf{y}_{1,2})(\mathbf{r}) 
	+ (\mathbf{x}_{1,2}*\mathbf{y}_{2,2})(\mathbf{r}),\\&
	\mathbf{z}_{2,1}(\mathbf{r}) = (\mathbf{x}_{2,1}*\mathbf{y}_{1,1})(\mathbf{r}) + (\mathbf{x}_{2,2}*\mathbf{y}_{2,1})(\mathbf{r}),  \;\; \\&
	\mathbf{z}_{2,2}(\r) = (\mathbf{x}_{2,1}*\mathbf{y}_{1,2})(\mathbf{r}) + (\mathbf{x}_{2,2}*\mathbf{y}_{2,2})(\mathbf{r}).
	\end{align*}	
	\item For a scalar image $g$ and an image of  $m\times l$ matrices, $\mathbf M$, ${\mathbf M}*g$ is defined as 
	$(\mathbf M*g)_{i,j}(\r)=({m}_{i,j}*g)(\r)$ for $i=1,...,m $, $j=1,...,n$ and any location $\r$.  
	\item For an  image of  $m\times l$ matrices, $\mathbf M$, we can define a new matrix image $\mathbf{N}=\mathbf{M}^t$ by extending the idea of transpose of a matrix to the case of matrix images. The $(i,j)$ entry of $\mathbf{N}$ can be defined as  $n_{i,j}(\r)=m_{j,i}(\r)$ for $i=1,...,m $, $j=1,...,n$ and any location $\r$. 
		\item
		For scalar images $u$ and $v$, and a scalar filter $h$,  the convolved inner product
	$\langle u, h*v\rangle$  satisfies the relation $\langle u, h*v\rangle= \langle \tilde{h}*u, v \rangle$,
	where $\tilde{h}$ denotes the flipped filter, i.e.,  $\tilde{h}$ satisfies $\tilde{h}({\mathbf r})=
	h(-{\mathbf r})$.  This relation can be easily verified by writing the inner product in Fourier domain.
	\item
	We will extend the notion of flipping for vector filter in a straight forward way.  In other words,  for a vector filter
	${\mathbf h}$,   $\tilde{\mathbf h}$ denotes the vector filter obtaining by flipping each of its  constituent scalar filters.
	For a vector image ${\mathbf u}$ and a vector filter ${\mathbf h}$,   and a scalar image of appropriate size, $v$, we have
	$\langle {\mathbf u},   \mathbf{h}*v\rangle = \langle \tilde{\mathbf{h}}^t*{\mathbf u},   v\rangle$. 	
	The operation $\tilde{\mathbf{h}}^t*(\cdot)$ is the adjoint of the operation $\mathbf{h}*(\cdot)$.
		\item
	Let ${\mathbf N}$ be an $m\times m$ matrix image,  and let ${\mathbf u}$  and ${\mathbf v}$ be
	$m\times 1$ vector filters. Then using the ideas of the previous point, we can show
	that $\langle {\mathbf u}, {\mathbf N}*\tilde{\mathbf v}\rangle = \langle {\mathbf u}*{\mathbf v}^t, {\mathbf N} \rangle$.  
	By a trivial extension of this relation,  we can also show that 
	$\langle {\mathbf N}, {\mathbf u}*{\mathbf v}^t*g \rangle = \langle {\mathbf N}*\tilde{\mathbf v},
	{\mathbf u}*g \rangle$  for any scalar image $g$.
	\item
	Let  ${\mathbf d}(\r)=
	[{d_{x}(\r)} \ \ {d_{y}(\r)}]^t$   where  $d_{x}(\r)$ and $d_{y}(\r)$ denote discrete  filters implementing
	the derivative operators
	$\frac{\partial}{\partial x}$ and  
	$\frac{\partial}{\partial y}$.  Then for a scalar image,  $g$,    $\mathbf{d}*g$   is the discrete gradient of $g$.
	Its adjoint,  
	$\tilde{\mathbf{d}}^t*(\cdot)$,  which is defined for image of $2\times 1$ vectors,  is called the discrete
	divergence. For a $2 \times 1$ vector image $\mathbf{v}$, we write  $div\; \mathbf{v} = 
	\tilde{\mathbf{d}}^t*\mathbf{v}$.  
	\item
	The extension of the notion of gradient for vector images is called the Jacobian.  For a   vector image,
	$\mathbf{v}(\r)$,  the Jacobian is given by  $(\mathbf{v}* \mathbf{d}^t)(\r)$.   For a $2\times 1$ vector image,  $\mathbf{v}(\r)$,
	 and $2\times 2$ matrix image, $\mathbf{U}(\r)$, the inner product $\langle  \mathbf{d}*\mathbf{v}^t,  \mathbf{U} \rangle$
	 satisfies  $\langle  \mathbf{d}*\mathbf{v}^t,  \mathbf{U} \rangle  =  \langle \mathbf{v}^t,  \mathbf{u} \rangle$
	 where $\mathbf{u}$ is the row vector image obtained by applying the adjoint of the operation  $\mathbf{d}*\mathbf{v}^t$
	 on $\mathbf{U}$.  $\mathbf{u}$ is given by  
	$div \; \mathbf{U} = [div\; \mathbf{u}_1 \; div\;\mathbf{u}_2]$.

\end{enumerate}
\section{Generalized Hessian-Schatten norm  regularization}

Let   $g(\mathbf{r})$ be the discrete candidate image, where ${\mathbf r}=[x\; y]^t$ is
the discrete pixel index. 
Recall that, ${\mathbf H}(\r)=
\genm{d_{xx}(\r)}{d_{yy}(\r)}{d_{xy}(\r)}$  denotes the  discrete Hessian operator. Convolution of this operator with   $g(\mathbf{r})$ (denoted  by
$({\mathbf H}*g)(\r)$)  is the discretized Hessian of the  candidate image. We regard ${\mathbf M}(\r)=({\mathbf H}*g)(\r)$ as an image of $2\times 2$
matrices;  in other words, for each pixel index, $\r$,   ${\mathbf M}(\r)$ is a $2\times 2$
matrix.  Using this formulation, the  well-known second-order total variation regularization
can  be expressed as $R_{TV2}(g)=\sum_{\r}\|({\mathbf H}*g)(\r)\|_F$.  Similarily, Hessian-Schatten norm regularization  \cite{lefkimmiatis2013hessian}
of order $p$ applied on the candidate image $g$  can be expressed as 
\begin{equation}
{\cal HS}_p(g, \alpha_s)= \alpha_s \|{\mathbf H}*g\|_{1,S(p)}  = \alpha_s \sum_{\r}\|({\mathbf H}*g)(\r)\|_{S(p)}.
\end{equation}
To develop the novel regularization by extending  Hessian-Schatten
norm,  we use the dual form of the Schatten norm.  Specifically,  
we write  Hessian-Schatten norm  on $g$ as 
\begin{align}
\label{eq:hsmaxfp}
{\cal HS}_p(g, \alpha_s)&= \max_{\|{\mathbf N}\|_{(\infty,S(q))}\le \alpha_s}\langle {\mathbf N}, {\mathbf H}*g \rangle.
\end{align}
Note that the above maximization is within the space of $2\times 2$ matrix images.  The maximizer
 will be a symmetric matrix image,  because    the Hessian,   ${\mathbf H}*g$,  is symmetric.  Hence,  
 ${\cal HS}_p(g, \alpha_s)$ will also be equal to the result of maximization  within the space of 
 $2\times 2$ symmetric  matrix images.  This is again equivalent to writing
 \begin{align}
{\cal HS}_p(g, \alpha_s)&= \max_{\|\overline{\mathbf N}\|_{(\infty,S(q))}\le \alpha_s}\langle {\overline{\mathbf N}}, {\mathbf H}*g \rangle.
\end{align}
To formulate the generalization, we write the above expression in expanded form given below:
\begin{equation}
{\cal HS}_p(g, \alpha_s)= \max_{\|\overline{\mathbf {N}}(\r)\|_{S(q)}\le \alpha_s}\sum_{\r} 
\langle \overline{\mathbf N}(\r), ({\mathbf H}*g)(\r) \rangle.
\end{equation}
From the above form, it is clear that the maximization is carried out for each ${\mathbf r}$ independently.
Suppose ${\cal B}_s(q)$ denotes the set of $2\times 2$  symmetric matrices such that,  
for each ${\mathbf A}\in {\cal B}_s(q)$  we have  $\|{\mathbf A}\|_{S(q)} \le 1$.  Then the above minimization
is carried out within a set that has a size that  is $L^2$ times  the size of ${\cal B}_s(q)$, where
$L\times L$  is the size of $g(\r)$. We propose to generalize ${\cal HS}_p(g, \alpha_s)$  by restricting
the maximization space in the same way the second order TGV generalizes the second order total variation
regularization (TV-2) \cite{tgv}.  To this end, let $\overline{\mathbf {N}}(\r)$ be of the form 
$\overline{\mathbf {N}}(\r)=\genm{n_{xx}(\r)}{n_{yy}(\r)}{n_{xy}(\r)}$, and  let 
$div\;(\cdot)$ for a matrix image be as defined in \cref{notation}.
Note that $(div\;\overline{\mathbf {N}})(\r)$  is vector image. 
With this, we express the generalization as given below:
\begin{align}
{\cal GHS}_p(g, \alpha_s, \alpha_f)&= 
\max_{\substack{\|\overline{\mathbf {N}}\|_{\infty,S(q)}\le \alpha_s\\
		\|div\;\overline{\mathbf {N}}\|_{\infty,2}\le \alpha_f} }\langle\overline{\mathbf {N}}, {\mathbf H}*g \rangle.
\end{align}
By using the fact that  $(div\;\overline{\mathbf {N}})=\tilde{\mathbf{d}}^t*\overline{\mathbf {N}}$
and transposition does not affect norm,
 we can write 
${\cal GHS}_p(g, \alpha_s, \alpha_f)$ as given below:
\begin{align}
		\label{eq:ghsfp}
{\cal GHS}_p(g, \alpha_s, \alpha_f)&= 
\max_{\substack{\|\overline{\mathbf {N}}\|_{\infty,S(q)}\le \alpha_s\\
		\|\overline{\mathbf {N}}*\tilde{\bf d}\|_{\infty,2}\le \alpha_f} }\langle {\overline{\mathbf N}}, {\mathbf H}*g \rangle .
\end{align}

Note that ${\cal GHS}_p(g, \alpha_s, \alpha_f)$  is a generalization in  the sense that
${\cal HS}_p(g, \alpha_s)$  is special case of 
${\cal GHS}_p(g, \alpha_s, \alpha_f)$, i.e., 
${\cal GHS}_p(g, \alpha_s, \infty)={\cal HS}_p(g, \alpha_s)$. 
The above form is not usable as a regularization for image recovery as it involves 
symmetrization operation.   Further, directly using the above form will lead to min-max
problem, which will not be very convenient for developing numerical minimization.
Interestingly, the above form can be translated into a form 
that appears very close the form TGV-2  given in the equation
\eqref{eq:tgv2imp} with the matrix Frobenius norm replaced by  the Schatten norm.
The following proposition gives  this expression.
\begin{prop}
	The generalized Hessian-Schatten norm ${\cal GHS}_p(g, \alpha_s, \alpha_f)$  can be expressed by
	\begin{align}
	{\cal GHS}_p(g, \alpha_s, \alpha_f) & = \underset{{\mathbf u}}{\operatorname{min}}\;\;
	\alpha_f \|{\mathbf d}*g - {\mathbf u}\|_{1,2}  + \alpha_s 
	\| 0.5({\mathbf u}*{\mathbf d}^t + {\mathbf d}*{\mathbf u}^{t})\|_{1,S(p)}.
	\end{align}
\end{prop}

Note that, in terms of pixel-wise summations, the above form of regularization can written 
as 
\begin{align}
\label{eq:ghs2}
{\cal GHS}_p(g, &\alpha_s, \alpha_f) = \underset{{\mathbf u}}{\operatorname{min}}\;\;
\alpha_f \sum_\r\|({\mathbf d}*g)(\r) - {\mathbf u}(\r)\|_{2} + \alpha_s \sum_r
\| 0.5(({\mathbf u}*{\mathbf d}^t)(\r) + ({\mathbf d}*{\mathbf u}^{t})(\r))\|_{S(p)}.
\end{align} 	
Note that this form resembles with  TGV-2 given in the equation	\eqref{eq:tgv2imp},  except the
fact that $\|\cdot\|_F$  is replaced by  $\|\cdot\|_{S(p)}$.  By noting the fact that 
$\|\cdot\|_{S(p)}$ with $p=2$ becomes the same as  $\|\cdot\|_F$, we conclude that the 
proposed regularization includes TGV-2 as a special case. 
Further note that ${\cal GHS}_p(g, \alpha_s, \alpha_f)$  becomes the Hessian-Schatten norm
regularization    as $\alpha_f \rightarrow \infty$.   Moreover, it becomes the TV-2
regularization with $p=2$ as $\alpha_f \rightarrow \infty$.
It can  be observed that the new regularization combines the best of Hessian-Schatten
norm regularization and TV-1 regularization, in the same way as TGV-2 combines the best of
TV-1 and TV-2 regularization.  
Near edges, ${\mathbf u}({\mathbf r})$ approaches zero leading to TV-1-like 
behaviour which allows sharp jumps in
the edges.  On the other hand,  in smoother regions,  ${\mathbf u}({\mathbf r})$ approaches  
${\mathbf d}*g({\mathbf r})$ leading
to the structure preserving ability of Hessian-Schatten norm.

\section{Image reconstruction algorithm using GHSN regularization}  

\subsection{The cost function}
\label{sec:costform}
In our development,  we will   restrict the  image measurement
model to be  in convolutional form.  We do this to gain the convenience
in implementing the forward model, as our main focus is on the 
regularization.   Let $h(\r)$
denote the impulse response of the imaging system and suppose that the
noise is Gaussian.   We allow the possibility that $h(\r)$ can be complex,
and hence the measurement can have complex values.
With this,   the data fitting  term  is given by  
$G(m,g) = \sum_{\r}\|(h*g)(\r) - m(\r)\|_{2}^2   = \|h*g - m\|_{2,2}^2$, 
and the 
GHS regularized
image reconstruction can be expressed as
\begin{equation}
g_{opt} = \underset{g}{\operatorname{argmin}} \;\; \|h*g - m\|_{2,2}^2 +
{\cal GHS}_p(g, \alpha_s, \alpha_f).
\end{equation}
By accounting for the fact that 
${\cal GHS}_p(g, \alpha_s, \alpha_f)$ itself is
expressed via a minimization,  we can also write the above problem as,
\begin{equation}
(g_{opt}, {\mathbf u}_{opt}) = \underset{g, {\mathbf u}}{\operatorname{argmin}} \;\; \|h*g - m\|_{2,2}^2 +
{\cal PGHS}_p(g, {\mathbf u},\alpha_s, \alpha_f),
\end{equation}
where 
\begin{align}
{\cal PGHS}_p(g,& {\mathbf u},\alpha_s, \alpha_f) =
\alpha_f \|{\mathbf d}*g - {\mathbf u}\|_{1,2} + \alpha_s
\| 0.5({\mathbf u}*{\mathbf d}^t + {\mathbf d}*{\mathbf u}^{t})\|_{1,S(p)}.
\end{align}
For notational convenience in developing the minimization algorithm,  we re-express the cost
in terms of the combined variable ${\mathbf v(\r)}=[g(\r)\;{\mathbf u}^t(\r)]^t$.  To this end, we
define the following:
\begin{align}
{\mathbf h}(\r) & = [h(\r)\;\;0\;\;0]^t,  \\
{\mathbf T}_f(\r) & = \left[\begin{array}{ccc}
d_x(\r) & 0 & 0 \\ 
d_y(\r) & 0 & 0 \\
0      & 1 & 0 \\
0      & 0 & 1 
\end{array}   \right],  \\
{\mathbf T}_s(\r) & = 	\left[ \begin{array}{ccc}
0 &     d_x(\r) & 0         \\ 
0 &  d_y(\r) & 0         \\
0 & 0       & d_x(\r)   \\ 
0 & 0       & d_y(\r)    
\end{array} \right]\text{, and}   \\
{\mathbf A}_f     & = \left[ \begin{array}{cccc}
1/\sqrt{2} & 0  & -1/\sqrt{2}  & 0 \\
0 & 1/\sqrt{2} & 0  & -1/\sqrt{2}  
\end{array}  \right].
\end{align}
We will overload the notation $\|\cdot \|_{S(p)}$,  such that when used with a
$4\times 1$ vector ${\mathbf y}=[y_1\; y_2\; y_3\;y_4]^t$ as the argument,
it represent the following: 
$$\|{\mathbf y}\|_{{S}(p)} =  \left\|  \left[
\begin{array}{cc}
y_1 & y_2 \\
y_3 & y_4   \end{array}
\right]\right\|_{S(p)}.$$
Let ${\mathbf K}$ be the matrix such that 
${\mathbf Ky} = [y_1\;\;0.5(y_2+y_3)\;\;0.5(y_2+y_3)\;\;y_4]^t$.
With these,  the overall cost to be minimized
can be expressed as
\begin{align}
\label{eq:costpghs}
J({\mathbf v}, &\alpha_s,\alpha_f, p) = \|{\mathbf h}*{\mathbf v} - m\|_{2,2}^2 +
{\alpha_f}{\sqrt{2}} \|{\mathbf A}_f({\mathbf T}_f*{\mathbf v})\|_{1,2} 
+ \alpha_s \|{\mathbf K}({\mathbf T}_s*{\mathbf v})\|_{1,{S(p)}}.
\end{align}
Note that definition of $\mathbf A_f$ creates an additional scale factor of $\sqrt{2}$ in the middle term in  \cref{eq:costpghs}. But, this does not create any difference in the formulation as $\alpha_f$ is an independent parameter, which is tuned for optimum reconstruction quality. Hence we absorb the scaling ${\sqrt{2}}$ into the parameter $\alpha_f$. Henceforth, the weight parameter  in the middle term will be $\alpha_f$ instead of ${\alpha_f}{\sqrt{2}}$.

Note that  $J({\mathbf v}, \alpha_s,\alpha_f, p)$  with $p=2$ is the well-known  TGV-2 regularization.    Guo et al.  \cite{guo2014new}   developed an algorithm for image reconstruction using TGV-2 and shearlet regularization.  With the 
 shearlet part removed,  their algorithm  corresponds to the following constrained formulation
of the reconstruction problem:
\begin{align}
({\mathbf v}^*, {\mathbf w}_s^*, {\mathbf w}_f^*)  & =
\underset{{\mathbf v}, {\mathbf w}_s, {\mathbf w}_f}
{\operatorname{argmin}}\;\; \|{\mathbf h}*{\mathbf v}-m\|_{2,2}^2 +
\alpha_f \|{\mathbf w}_f\|_{1,2}  + \alpha_s \|
{\mathbf w}_s\|_{1,{S(2)}}  \\
\nonumber
& \mbox{subject to}\;\; {\mathbf w}_s = {\mathbf K}({\mathbf T}_s*{\mathbf v}),
\;\;{\mathbf w}_f = {\mathbf A}_f({\mathbf T}_f*{\mathbf v}).
\end{align}
The main advantage of this form is that the ADMM algorithm can be constructed using well known proximal
operators.  However, the algorithm can be badly conditioned in  some cases as we will demonstrate in the
experiment section. 
 	
\subsection{Proposed ADMM  algorithm}
	
	Before specifying the variable splitting scheme for the proposed ADMM algorithm,  we first rewrite the
	cost of the equation \eqref{eq:costpghs} as
	\begin{align}
	\label{eq:costbghs2}
J({\mathbf v}, \alpha_s,\alpha_f, p)& = \|{\mathbf h}*{\mathbf v} - m\|_{2,2}^2 + 
	\alpha_f \|{\mathbf A}_f({\mathbf T}_f*{\mathbf v})\|_{1,2} 
	+ \alpha_s \|{\mathbf K}({\mathbf T}_s*{\mathbf v})\|_{1,{S(p)}} + {\cal B}(\mathbf e^t*\mathbf v),
	\end{align}
	where  ${\cal B}(g)$ is the bound  constraint function, which takes infinity when any of the pixel of
	$g$ is outside the user-defined bound   and zero otherwise, and ${\mathbf e}^t(\r) = [\delta(\mathbf r)\;0\;0]$ with $\delta(\mathbf r)$ denoting Kronecker delta, and hence
	${\mathbf e}^t*{\mathbf v} = v_1 = g$.
	We propose to build ADMM algorithm by means of the following constrained formulation:
	\begin{align}
         ({\mathbf v}^*, {\mathbf w}_s^*, {\mathbf w}_f^*, {w}_b^*) = &
	\underset{{\mathbf v}, {\mathbf w}_s, {\mathbf w}_f, {w}_b}{\operatorname{argmin}}\;\; \|{\mathbf h}*{\mathbf v}-m\|_2^2 +
	\alpha_f \|{\mathbf A}_f{\mathbf w}_f\|_{1,2}  + 
	\alpha_s \|{\mathbf K \mathbf w}_s\|_{1,{S(p)}} + {\cal B}(w_b) \\
	\nonumber
	& \mbox{subject to}\;\; {\mathbf w}_s = {\mathbf T}_s*{\mathbf v}, \;\;{\mathbf w}_f = {\mathbf T}_f*{\mathbf v},
	\;\;{\mathbf e}^t*{\mathbf v} = w_b.
	\end{align}
	 The main difference from the previous
	formulation \cite{guo2014new} is that the matrices ${\mathbf A}_f$  and  ${\mathbf K}$ are not a part of the constraints, but,
	are left as a part of the cost to be minimized.  This leads to some numerical advantages, which
	we will clarify  after completing the development of the algorithm.  On the other hand, this splitting
	scheme requires constructing new proximal operators for implementing the ADMM algorithm.  We do this
	in the next section which is  one of the important contributions of the paper. In the remainder of this section,
	we complete specifying the ADMM iteration for splitting scheme specified above.
	In order to facilitate developing  algorithm  in terms of compact expressions, we need to further simplify the
	notations. To this
	end,  let 
	\begin{align}
	\label{eq:wdef}
	{\mathbf w}  & = [w_b\; {\mathbf w}_s^t\; {\mathbf w}_f^t]^t  \text { and }\\ 
	\label{eq:tdef}
	{\mathbf T}(\r) & = \left[ \begin{array}{c}
	{\mathbf e^t}(\r) \\
	{\mathbf T}_s(\r) \\
	{\mathbf T}_f(\r) 
	\end{array} 
	\right].
	\end{align}
	With this, the above problem can be expressed as
	\begin{align}
	({\mathbf v}^*, {\mathbf w^*}) = & 
	\underset{{\mathbf v}, {\mathbf w}}{\operatorname{argmin}}\;\; \|{\mathbf h}*{\mathbf v}-m\|_2^2 +
	R({\mathbf w},\alpha_s,\alpha_f,p) \\
	& \mbox{subject to}\;\; {\mathbf w} = {\mathbf T}*{\mathbf v},
	\end{align}
	where 
	\begin{equation}
	R({\mathbf w},\alpha_s,\alpha_f,p) = \alpha_f \|{\mathbf A}_f{\mathbf w}_f\|_{1,2}  + 
	\alpha_s \|{\mathbf K \mathbf  w}_s\|_{1,{S(p})} + {\cal B}(w_b).
	\end{equation}
	The next step towards developing the ADMM algorithm is to write the augmented
	Lagrangian of the above constrained optimization problem.  The augmented Lagrangian
	is given by  
	\begin{align}
	L({\mathbf v}, {\mathbf w}, {\pmb \lambda}, \alpha_s,\alpha_f, p ) &=
	\|{\mathbf h}*{\mathbf v}-m\|_2^2 +
	R({\mathbf w},\alpha_s,\alpha_f,p) + 
	\frac{\beta}{2} \|{\mathbf T}*{\mathbf v}-{\mathbf w} \|_2^2 +
	\langle {\pmb \lambda}, {\mathbf T}*{\mathbf v}-{\mathbf w} \rangle,
	\end{align}
	where ${\pmb \lambda}$ is vector image of Lagrange's multiplier with 
	its dimension equal to that
	of ${\mathbf w}$. Also, $\beta$ is a fixed positive real number.  The ADMM becomes series of minimizations with respect 
	to ${\mathbf v}$ and ${\mathbf w}$ and
	updates on ${\pmb \lambda}$.  
	Given the current set of iterates, 
	$\{{\mathbf v}^{(k)}, {\mathbf w}^{(k)}, {\pmb \lambda}^{(k)}\}$  the ADMM 
	methods proceeds as follows:
	\begin{align}
	\label{eq:admmwp}
	{\mathbf w}^{(k+1)} & = \underset{{\mathbf w}}{\operatorname{argmin}}\;\;
	L({\mathbf v}^{(k)}, {\mathbf w}, {\pmb \lambda}^{(k)}, \alpha_s,\alpha_f, p), \\
	\label{eq:admmvp}
	{\mathbf v}^{(k+1)} & = \underset{{\mathbf v}}{\operatorname{argmin}}\;\;
	L({\mathbf v}, {\mathbf w}^{(k+1)}, {\pmb \lambda}^{(k)}, \alpha_s,\alpha_f, p), \\
	\label{eq:admmyp}
	\text {and }{\pmb \lambda}^{(k+1)} & = 
	{\pmb \lambda}^{(k)} + \beta \left({\mathbf T}*{\mathbf v}^{(k+1)}-{\mathbf w}^{(k+1)} \right).
	\end{align}
	
\section{Solving the sub-problems of ADMM}

\subsection{The ${\mathbf w}$-problem}

\subsubsection{{Expressing the pixel-wise sub-problems}}

Note that the solution to the minimization problem of equation \eqref{eq:admmwp}, 
${\mathbf w}^{(k+1)}$,   is also the minimum of the following cost:
\begin{equation}
\label{eq:wp2}
L_{w,k}({\mathbf w},\alpha_s,\alpha_f, p) = R({\mathbf w},\alpha_s,\alpha_f,p) + \frac{\beta}{2}
\|{\mathbf w}-{\mathbf x}^{(k)}\|_2^2,
\end{equation}
where, 
\begin{equation}
\label{eq:wkbar}
\bar{\mathbf x}^{(k)} = {\mathbf T}*{\mathbf v}^{(k)} + \frac{1}{\beta}{\pmb \lambda}^{(k)}.
\end{equation}
For notational convenience,  let ${\mathbf x}={\mathbf x}^{(k)}$ and let
$\hat{\mathbf w}= {\mathbf w}^{(k+1)}$.   Note that the cost $L_{w,k}({\mathbf w},\alpha_s,\alpha_f, p)$
is separable  across the subvectors of the variable, ${\mathbf w}$; in other words, it is separable
across the subvectors ${w}_b$, ${\mathbf w}_s$, and ${\mathbf w}_f$, and we introduced the collective
variable, ${\mathbf w}$ only for notational convenience in expressing the  ADMM loop of equations
\eqref{eq:admmwp}, \eqref{eq:admmvp}, \eqref{eq:admmyp}. As we are focused on this specific
sub-problem,  we will now separate  the constituent problems. To this end, 
let $x_b$, ${\mathbf x}_s$, and ${\mathbf x}_f$
denote the sub-vectors of ${\mathbf x}$ conferring  to the partitioning given in the
equation \eqref{eq:wdef}. Similarly, let $\hat{w}_b$, $\hat{\mathbf w}_s$, and $\hat{\mathbf w}_f$
be the subvectors of  $\hat{\mathbf w}$. 
With this the solution to the ${\mathbf w}$-problem can be expressed as
\begin{align}
\label{eq:wbprob2}
\mbox{$w_b$-problem:} \;\;\;&  \hat{w}_b = 
\underset{w_b}{\operatorname{argmin}} \;\; 
\underbrace{\frac{\beta}{2} \|x_b-w_b\|_{2,1}^2 + 
	{\cal B}(w_b)}_{\bar{L}_{b}(w_b,x_b)}  \\
\label{eq:wsprob2}
 \mbox{${\mathbf w}_s$-problem:} \;\;\; &  \hat{\mathbf w}_s = 
\underset{{\mathbf w}_s}{\operatorname{argmin}} \;\; 
\underbrace{\frac{\beta}{2} \|{\mathbf x}_s-{\mathbf w}_s\|_{2,2}^2 + 
	\alpha_s \|\mathbf{K w}_s\|_{1,{S(p)}}}_{\bar{L}_{s}({\mathbf w}_s, 
	{\mathbf x}_s, \alpha_s)} \\
\label{eq:wfprob2}
\mbox{${\mathbf w}_f$-problem:} \;\;\; &  \hat{\mathbf w}_f = 
 \underset{{\mathbf w}_f}{\operatorname{argmin}} \;\; 
\underbrace{\frac{\beta}{2} \|{\mathbf x}_f-{\mathbf w}_f\|_{2,2}^2 + 
	\alpha_s \|{\mathbf A}_f\mathbf{w}_f\|_{1,2}}_{\bar{L}_{f}({\mathbf w}_f, {\mathbf x}_f, \alpha_f)}
\end{align}
Note that the functions $\bar{L}_{b}(w_b,x_b)$, 
$\bar{L}_{s}({\mathbf w}_s, {\mathbf x}_s, \alpha_s)$, and 
$\bar{L}_{f}({\mathbf w}_f, {\mathbf x}_f, \alpha_f)$  are separable across pixel indices since 
they are constructed as mixed norms composed elementary pixel-wise norms. 
Hence, they can be expressed as a sum of pixel-wise elementary functions. 
The form of these elementary functions can be clearly deduced from the form of
the functions $\bar{L}_{b}(w_b,x_b)$, 
$\bar{L}_{s}({\mathbf w}_s, {\mathbf x}_s, \alpha_s)$, and 
$\bar{L}_{f}({\mathbf w}_f, {\mathbf x}_f, \alpha_f)$.
We can express these functions as
\begin{align}
\label{eq:wbprob3}
\bar{L}_{b}(w_b,x_b)   & = \sum_\r 
\underbrace{\frac{\beta}{2} (x_b(\r)-w_b(\r))^2 + 
	{\cal B}(w_b(\r))}_{
	{L}_{b}(w_b(\r),x_b(\r)) }, \\
 \label{eq:wsprob3}
\bar{L}_{s}({\mathbf w}_s, {\mathbf x}_s, \alpha_s)  & = \sum_\r
\underbrace{\frac{\beta}{2}\|{\mathbf x}_s(\r)-{\mathbf w}_s(\r)\|_{2}^2 + 
	\alpha_s \|{\mathbf K w}_s(\r)\|_{{S(p)}}}_{
	{L}_{s}({\mathbf w}_s(\r), {\mathbf x}_s(\r), \alpha_s)},
\\
\label{eq:wfprob3}
\bar{L}_{f}({\mathbf w}_f, {\mathbf x}_f, \alpha_f)  & = \sum_\r
\underbrace{
	\frac{\beta}{2} \|{\mathbf x}_f(\r)-{\mathbf w}_f(\r)\|_{2}^2 + 
	\alpha_f \|{\mathbf A}_f({\mathbf w}_f(\r))\|_{2}}_{
	{L}_{f}({\mathbf w}_f(\r), {\mathbf x}_f(\r), \alpha_f)  }.
\end{align}
From the above form, it is clear that the minimization problems 
are separable across pixel indices. Hence the solution to the
minimization problems of equations \eqref{eq:wbprob2}, \eqref{eq:wsprob2},
and \eqref{eq:wfprob2}, can be expressed as the following:
\begin{align}
\label{eq:wbrobpix}
\hat{w}_b(\r) = & 
\underset{z\in \mathbb{R}}{\operatorname{argmin}} \;\;
{L}_{b}(z,x_b(\r)), \\
\label{eq:wsprobpix}
\hat{\mathbf w}_s(\r) = &
\underset{{\mathbf z}\in \mathbb{R}^4}{\operatorname{argmin}} \;\;
{L}_{s}({\mathbf z},{\mathbf x}_s(\r), \alpha_s)\text{ , and}  \\
\label{eq:wfprobpix}
\hat{\mathbf w}_f(\r) = &
\underset{{\mathbf z}\in \mathbb{R}^4}{\operatorname{argmin}} \;\;
{L}_{f}({\mathbf z},{\mathbf x}_f(\r), \alpha_f).
\end{align}

\subsubsection{{Solution to  the pixel-wise sub-problems}}
\label{sec:pwsubp}

The solution to the
$w_b$-problem  is very simple, and it is the clipping of the pixels by bound
that defines ${\cal B}(w_b)$  \cite{parikh2014proximal}.
We express the solution as given below:
\begin{equation}
\hat{w}_b(\r) =  \mathbb{P}_{ b}(x_b(\r)),
\end{equation}
where ${\mathbb{ P}}_{b}(\cdot)$  denotes the operation of clipping the pixel values within
the specified bounds.
Next, we consider expressing the solution for  ${\mathbf w}_f$-problem.  Note that, in the
absence of the matrix ${\mathbf A}_f$, the solution will be the well-known shrinkage operation
on the $l_2$ norm of ${\mathbf x}_f(\r)$ \cite{parikh2014proximal}.  Because of the presence of the matrix, 
${\mathbf x}_f(\r)$,
the shrinkage operation is not directly applicable.  The following proposition gives the 
solution   to the ${\mathbf w}(\r)$ problem.
\begin{prop}
	The minimum of $L_f({\mathbf z}, {\mathbf a}, \alpha_f)= \frac{\beta}{2}\|{\mathbf z}-{\mathbf a}\|_2^2  + 
	\alpha_f \|{\mathbf A}_f{\mathbf z}\|_2$
	with respect to ${\mathbf z}$ is given by ${\cal P}({\mathbf a}, t)={\mathbf a}-
	\min(\|{\mathbf A}_f{\mathbf a}\|_2, t){\mathbf A}_f^t{\mathbf A}_f{\mathbf a}$,  where
	$t=\alpha_f/\beta$. 
\end{prop}  
Now,  considering the  ${\mathbf w}_s$ problem,  i.e.,  considering the minimization of 
$\frac{\beta}{2}\|{\mathbf z}-{\mathbf a}\|_2^2  + 
\alpha_s \|{\mathbf K z}\|_{{S(p)}}$, the presence of ${\mathbf K}$  makes the problem more complex.
Otherwise, the solution to the problem in the absence of ${\mathbf K}$ is well-known \cite{poisson_hessian}.
The following proposition gives
expression for solution to this problem.
\begin{prop}
	The minimum of 
	$L_s({\mathbf z}, {\mathbf a}, \alpha_s)= \frac{1}{2}\|{\mathbf z}-{\mathbf a}\|_2^2  + 
	t \|{\mathbf K z}\|_{{S(p)}}$  with respect to ${\mathbf z}$ is given by
	$\bar{\cal P}_s({\mathbf a},t) = {\mathbf L}{\mathbf a} + {\cal P}_s({\mathbf K}{\mathbf a},t)$
	where ${\cal P}_s({\mathbf b},t)$  represents the minimum of the 
	$\frac{1}{2}\|{\mathbf z}-{\mathbf b}\|_2^2  + 
	t \|{\mathbf z}\|_{{S(p)}}$ with respect to ${\mathbf z}$,  and
	and ${\mathbf L}$  is the matrix such that
	its operations on ${\mathbf a}=[a_1\;a_2\;a_3\;a_4]^t$ is defined as
	${\mathbf L\a} = [0\;\;0.5(a_2-a_3)\;\;0.5(a_3-a_2)\;\;0]^t$.
\end{prop}
Note that ${\cal P}_s(\cdot,t)$  is the well-known proximal operator of 
Schatten norm (for details see  \cite{poisson_hessian}), and the above proposition expresses the required proximal
operator---proximal operator for the modified Schatten norm,
$\bar{\cal P}_s(\cdot,t)$---as a simple modification of ${\cal P}_s(\cdot,t)$.
Note that we just need set $t=\alpha_s/\beta$  for applying the above proposition
for the problem of equation \eqref{eq:wsprobpix}. Although  ${\cal P}_s(\cdot,t)$
is well known, 
in {\bf Algorithm 3},  we provide detailed description of  ${\cal P}_s(\cdot,t)$
to the level required for implementation along with a self-contained description
of the overall algorithm.  Note that, we have non-iterative exact formula for
${\cal P}_s(\cdot,t)$  only for $p=1,2$.  Hence, in the experimental 
demonstration,  we only consider these two values for $p$.

\subsection{The ${\mathbf v}$-problem}

Now we consider the minimization problem of equation \eqref{eq:admmvp}. 
By taking the dependencies of the minimization variable ${\mathbf v}$,  
it can be deduced that, the solution to this problem is also the minimum
of the following cost function:
\begin{equation}
\label{eq:admmvp3}
L_{v,k}({\mathbf v}) =
\|{\mathbf h}*{\mathbf v}-m\|_2^2 +
\frac{\beta}{2} \|{\mathbf T}*{\mathbf v}-{\mathbf y}^{(k)} \|_2^2,
\end{equation}
where
\begin{equation}
{\mathbf y}^{(k)} = {\mathbf w}^{(k+1)}-\frac{1}{\beta}{\pmb \lambda}^{(k)}.
\end{equation}
As done before for notational convenience, we let
${\mathbf y}={\mathbf y}^{(k)}$,  and $\hat{\mathbf v}={\mathbf v}^{(k+1)}$. 
Let ${\mathbf v}(\r) = [v_1(\r)\;v_2(\r)\;v_3(\r)]^t$, 
$\hat{\mathbf v}(\r) = [\hat{v}_1(\r)\;\hat{v}_2(\r)\;\hat{v}_3(\r)]^t$.
From the definition
of ${\mathbf T}(\r)$ given in the equation \eqref{eq:tdef}, 
and from the definition of  
${\mathbf h}(\r)$,
it is clear that the cost $L_{v,k}({\mathbf v})$ is separable across the
components of ${\mathbf v}(\r)$, which are 
$v_1(\r)$, $v_2(\r)$, and $v_3(\r)$.
They are given below:
\begin{align}
 \bar{L}_1(v_1)  & = \|h*v_1 - m \|_{2,2}^2 + 
\frac{\beta}{2} \|v_1 - y_b\|_{2,2}^2 
+ \frac{\beta}{2} \|d_x*v_1 - y_{f,1}\|_{2,2}^2 
+  \frac{\beta}{2} \|d_y*v_1 - y_{f,2}\|_{2,2}^2, \\
\bar{L}_2(v_2) & = 
\frac{\beta}{2} \|v_2 - y_{f,3}\|_{2,2} ^2
+ \frac{\beta}{2} \|d_x*v_2 - y_{s,1}\|_{2,2}^2 
+\frac{\beta}{2} \|d_y*v_2 - y_{s,2}\|_{2,2} ^2\text{, and} \\
\bar{L}_3(v_3) & = 
\frac{\beta}{2} \|v_3 - y_{f,4}\|_{2,2} ^2
 +\frac{\beta}{2} \|d_x*v_3 - y_{s,3}\|_{2,2}^2 
+ \frac{\beta}{2} \|d_y*v_3 - y_{s,4}\|_{2,2}^2.
\end{align}
Here,  ${\mathbf y}(\r) =[y_b(\r)\; 
y_{f,1}(\r)\;  y_{f,2}(\r)\;
y_{f,3}(\r)\;  y_{f,4}(\r)\;
y_{s,1}(\r)\;  y_{s,2}(\r)\;
y_{s,3}(\r)\;  y_{s,4}(\r)]^t$. 
These are quadratic functions involving simple discrete filtering operation. 
Hence, their minima can be expressed in terms of simple Fourier inversion.
To this end, we write the expression for the gradients below:
\begin{align}
\nabla_{v_1}\bar{L}_1(v_1)  ~=~ &   2 Re(conj(\tilde{h})*h)*v_1 +
 \beta(v_1+\tilde{d_x}*d_x*v_1 + \tilde{d_y}*d_y*v_1)  
   -2 Re(conj(\tilde{h})*m) \\
  \nonumber  & - \beta(y_b+\tilde{d}_x*y_{f,1} + \tilde{d}_y*y_{f,2}), \\
\nabla_{v_2}\bar{L}_2(v_2)  ~=~ &  
\beta(v_2+\tilde{d_x}*d_x*v_2 + \tilde{d_y}*d_y*v_2)  
  -\beta(y_{f,3}+\tilde{d}_x*y_{s,1} + \tilde{d}_y*y_{s,2})\text{, and} \\
\nabla_{v_3}\bar{L}_3(v_3)  ~=~ &  
\beta(v_3+\tilde{d_x}*d_x*v_3 + \tilde{d_y}*d_y*v_3)  
  -\beta(y_{f,4}+\tilde{d}_x*y_{s,3} + \tilde{d}_y*y_{s,4}). 
\end{align}
In the above equation, $conj(\cdot)$ represents pointwise complex conjugate operation.
Now, the minima $\hat{v}_1(\r)$, $\hat{v}_2(\r)$, and $\hat{v}_3(\r)$
can be obtained by solving $\nabla_{v_1}\bar{L}_1(v_1) = 0$,     
$\nabla_{v_2}\bar{L}_2(v_2)=0$, and   $\nabla_{v_3}\bar{L}_3(v_3)=0$
in Fourier domain. 
This is the main advantage of the proposed variable splitting:
we are able to solve for   $\hat{v}_1(\r)$, $\hat{v}_2(\r)$, and $\hat{v}_3(\r)$
independently by simple Fourier division.  On the other hand,
the ${\mathbf v}$-problem encountered in the splitting proposed in \cite{guo2014new}, 
the components $v_1$, $v_2$, and $v_3$ are coupled,  and hence, it requires
solving $3\times 3$ system of equations each of 
$L\times L$ frequencies values where $L\times L$ is the size of the
image to be reconstructed. The linear system to be solved in the $\mathbf{v}$ sub-problem do
 not have a block diagonal structure as obtained in our version of ADMM algorithm. Instead we 
 need to solve a non-diagonal block circulant system.
Although the required inverses can be
precomputed outside the ADMM loop, it turns out that overall 
numerical conditioning is worse than the splitting that we propose here,
which we demonstrate experimentally in Section  \ref{sec:exp}.
In   {\bf Algorithm 1},  we provide detailed expression of the solutions
of $\nabla_{v_1}\bar{L}_1(v_1) = 0$, 
$\nabla_{v_2}\bar{L}_2(v_2)=0$, and   $\nabla_{v_3}\bar{L}_3(v_3)=0$
to the level required for implementation. {\bf Algorithm 1} also serves
as the self-contained specification of the overall proposed reconstruction
method.
\begin{algorithm}[]
	\DontPrintSemicolon
	\SetAlgoLined
	Inputs: \\
	$\;\;\;\;m$:  $L\times L$  measured image \\
	$\;\;\;\;h$:  Impulse response of imaging system  \\
	$\;\;\;\;\alpha_f$:  regularization weight for first order term  \\
	$\;\;\;\;\alpha_s  $:  regularization weight for second order term  \\
	$\;\;\;\;\epsilon  $: termination tolerance  \\
	 Symbol definitions:  $d_x$  and $d_y$  are  first derivative filters\\
	$\;\;\;\;\;\;\;\;\;\;\;\;\;\;\;\;\;\;\;\;\;\;\;\;\;\;\;\;\;$ $\tilde{d}_x$  and $\tilde{d}_y$ are flipped versions \\
	$\;\;\;\;\;\;\;\;\;\;\;\;\;\;\;\;\;\;\;\;\;\;\;\;\;\;\;\;\;\;
	\;$  $\tilde{h}$ is flipped version of $h$\\
	\rule{.450\textwidth}{1pt}\\
	$\textbf{Initialization}:$\\
	$\;\;\;\;k \gets 0, \;   r_k \gets 1+\epsilon$ \\
	$\;\;\;\;{\mathbf v}^{(k)} \gets {\mathbf 0}_{3\times N \times N}$ \\
	$\;\;\;\;{\lambda }^{(k)}_b \gets {\mathbf 0}_{N \times N}$,
	${\pmb \lambda }^{(k)}_f \gets {\mathbf 0}_{4\times N \times N}$,
	${\pmb \lambda }^{(k)}_s \gets {\mathbf 0}_{4\times N \times N}$ \\
	$\;\;\;\;\bar{x}_b^{(k)} \gets  {\mathbf 0}_{N \times N}$ \\
	$\;\;\;\;\bar{\mathbf x}_f^{(k)} \gets {\mathbf 0}_{4\times N \times N} $ \\
	$\;\;\;\;\bar{\mathbf x}_s^{(k)} \gets  {\mathbf 0}_{4\times N \times N}$ \\
	$\;\;\;\; Q_d \gets  {\mathcal F \{Re(h*conj(\tilde{h}))}\}$ \\
	$\;\;\;\; Q_r \gets   1 + {\cal F}(d_x*\tilde{d}_x + d_x*\tilde{d}_x)$ \\
	$\;\;\;\; B_d \gets  {\mathcal F \{Re(conj(\tilde{h})*m)}\}$ \\
	\While{$r_k  \geq \epsilon$} 
	{   
		$\;\;\;~~~~~~~$~~~~\\
		$x_b^{(k)} \gets \bar{x}_b^{(k)}  + (1/\beta){\lambda }^{(k)}_b$ \\
		${\mathbf x}_f^{(k)} \gets \bar{\mathbf x}_f^{(k)} +  (1/\beta){\pmb \lambda }^{(k)}_f$ \\
		${\mathbf x}_s^{(k)} \gets\bar{\mathbf x}_s^{(k)}  +  (1/\beta){\pmb \lambda }^{(k)}_s$ \\
		~~~\\
		${w}^{(k+1)}_b \gets \mathbb{P}_b({x}_b^{(k)})~~~~~~~~~~~~~~$
		(Clipping of elements)\\
		$({\mathbf w}^{(k+1)}_f, {\mathbf w}^{(k+1)}_s) \gets 
		\mbox{\large Prox}( {\mathbf{x}}_f^{(k)}, {\mathbf{x}}_s^{(k)}, \alpha_f, \alpha_s)$  \;\;({\bf Alg. 2})\\
		$y_b^{(k+1)} \gets w_b^{(k+1)} - (1/\beta)\lambda_b^{(k)}$ \\
		${\mathbf{y}}_f^{(k+1)} \gets {\mathbf{w}}_f^{(k+1)} - (1/\beta){\pmb \lambda}_f^{(k)}$ \\
		${\mathbf{y}}_s^{(k+1)} \gets {\mathbf{w}}_s^{(k+1)} - (1/\beta){\pmb \lambda}_s^{(k)}$ \\
		~~~\\
		$B_1 \gets {\cal F}(y_b^{(k+1)}+\tilde{d}_x*y_{f,1}^{(k+1)} + \tilde{d}_y*y_{f,2}^{(k+1)})$  \\
		$B_2 \gets {\cal F}(y_{f,3}^{(k+1)}+\tilde{d}_x*y_{s,1}^{(k+1)} + \tilde{d}_y*y_{s,2}^{(k+1)})$ \\
		$B_3 \gets {\cal F}(y_{f,4}^{(k+1)}+\tilde{d}_x*y_{s,3}^{(k+1)} + \tilde{d}_y*y_{s,4}^{(k+1)})$ \\
		${v}^{(k+1)}_1 \gets   {\cal F}^{-1}((B_d+\beta B_1)/(Q_d + \beta Q_r))$ \\
		${v}^{(k+1)}_2 \gets   {\cal F}^{-1}(B_2/Q_r)$ \\
		${v}^{(k+1)}_3 \gets   {\cal F}^{-1}(B_3/Q_r)$ \\
		~~~\\
		$\bar{x}_b^{(k+1)} \gets v_1^{(k+1)}$ \\
		$\bar{\mathbf{x}}_f^{(k+1)} \gets [v_1^{(k+1)}*d_x, v_1^{(k+1)}*d_y, v_2^{(k+1)}, v_3^{(k+1)}]^t $ \\
		$\bar{\mathbf{x}}_s^{(k+1)} \gets [v_2^{(k+1)}*d_x, v_2^{(k+1)}*d_y, v_3^{(k+1)}*d_x, v_3^{(k+1)}*d_y]^t$ \\
		~~~\\
		$\lambda^{(k+1)}_b \gets \lambda^{(k)}_b + \beta ( \bar{x}_b^{(k+1)} - {w}^{(k+1)}_b)$ \\
		${\pmb \lambda}^{(k+1)}_f \gets {\pmb \lambda}^{(k)}_f + \beta ( \bar{\mathbf{x} }_f^{(k+1)} - {\mathbf{w}}^{(k+1)}_f)$ \\
		${\pmb \lambda}^{(k+1)}_s \gets {\pmb \lambda}^{(k)}_s + \beta ( \bar{\mathbf{x}}_s^{(k+1)} - {\mathbf{w}}^{(k+1)}_s)$ \\
		$k \gets k+1 $ 
	}
	\Return $\bf{y} = \textbf{x}^{(k)}$
	\caption{GHSN regularized Reconstruction:  
		$GHSN\text{-}RR(m, h, \alpha_0, \alpha_1, \epsilon)$}
	\label{alg:RR}
\end{algorithm}
\begin{algorithm}[]
	\DontPrintSemicolon
	\SetAlgoLined
	Inputs: \\
	$\;\;\;\;{\mathbf a}=[a_1\;a_2\;a_3\; a_4]^t$:  $4\times 1$  vector image corresponding to first order derivative \\
	$\;\;\;\;{\mathbf b}=[b_1\;b_2\;b_3\; b_4]^t$:  $4\times 1$  vector image corresponding to second order derivative \\
	$\;\;\;\;\alpha_f  $:  regularization weight for second order term  \\
	$\;\;\;\;\alpha_s  $:  regularization weight for second order term  \\
	\rule{.45\textwidth}{1pt}\\
	~~~\\
	\For {$\r \in [1,N]^2$} {  
		~~~\\
		$d_1(\r) \gets a_1(\r) - a_3(\r), \;\;\;  d_2(\r) \gets a_2(\r) - a_4(\r)$ \\
		$n(\r) \gets \sqrt{0.5(d_1^2(\r) + d_2^2(\r))}$ \\
		$n(\r) \gets 0.5\min (1, \alpha_f/n(\r))$ \\
		$d_1(\r) \gets n(\r)d_1(\r),\;\;\; d_2(\r) \gets n(\r)d_2(\r)$\\
		$\hat{\mathbf a}(\r) \gets [a_1(\r) - d_1(\r), a_3(\r) + d_1(\r),
		a_2(\r) - d_2(\r) ,  a_4(\r) + d_2(\r)]^t$\\
		~~~~\\
		$(b_1(\r),c(\r),b_4(\r)) \gets\mbox{ProxHS}
		(b_1(\r), 0.5(b_2(\r)+b_3(\r)), b_4(\r), \alpha_s, p)$ \\
		$d(\r) \gets 0.5(b_2(\r)-b_3(\r))$ \\
		$\hat{\mathbf b}(\r) \gets 
		[b_1(\r)\;\; c(\r) + d(\r) \;\; c(\r) -d(\r) \;\; b_4(\r)]^t$ \\
	}
	\mbox{Return}\;\;  $\hat{\mathbf a}, \hat{\mathbf b}$ \\
	
	\caption{Proximal operations: 	\\
		$(\hat{\mathbf a}, \hat{\mathbf c}) \gets 
		\mbox{\large Prox}({\mathbf a}, {\mathbf b}, \alpha_f, \alpha_s, p)$}
	\label{alg:proxa}
\end{algorithm}
\begin{algorithm}[]
	\DontPrintSemicolon
	\SetAlgoLined
	\If{p = 2} {
		$n \gets \sqrt{a^2+b^2+2c^2}$ \\
		$n \gets max(n-\alpha_s,0)/n$ \\
		$(\hat{a},\hat{b}, \hat{c}) \gets (na, nb, nc)$
	}
	\ElseIf{p = 1} 
	{
		$l_1 \gets (a+b) + \sqrt{(a-b)^2 + 4c}$ \\
		$l_2 \gets (a+b) - \sqrt{(a-b)^2 + 4c}$ \\
		${\mathbf v}_1 \gets [a-l_1\;\;c]^t$ \\
		${\mathbf v}_1 \gets {\mathbf v}_1/\|{\mathbf v}_1\|_2$ \\
		${\mathbf v}_2 \gets [a-l_2\;\;c]^t$ \\
		${\mathbf v}_2 \gets {\mathbf v}_2/\|{\mathbf v}_2\|_2$ \\
		$l_1	\gets  sign(l_1)max(abs(l_1)-\alpha_s,0)$\\
		$l_2	\gets  sign(l_2)max(abs(l_2)-\alpha_s,0)$\\
		$\hat{a} \gets v_{11}^2l_1 + v_{21}^2l_2 $\\
		$\hat{b} \gets v_{12}^2l_1 + v_{22}^2l_2 $\\

		$\hat{c} \gets v_{11}v_{12}l_1 + v_{21}v_{22}l_2$ 
	}
	\caption{Proximal operator for Hessian-Schatten norm: 	
		$(\hat{a}, \hat{c},  \hat{b}) \gets 
		\mbox{\large ProxHS}(a, c, b, \alpha_s, p)$}
	\label{alg:prox}
\end{algorithm}

\section{Experiments} \label{sec:exp}

\begin{figure*}[h!] 
	\caption{Images used in experiments}
	\centering
	\includegraphics[width=1\textwidth]{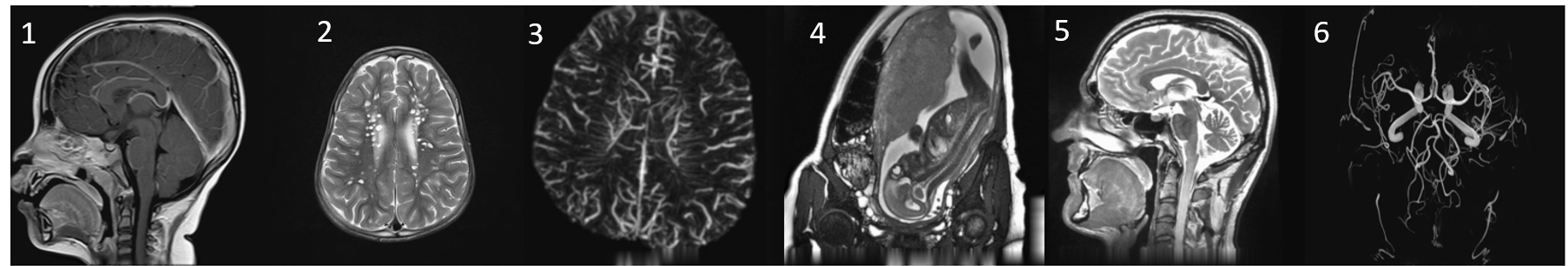}
	\label{Image: All Images}
\end{figure*}

\begin{figure}[h!] 
 	\caption{MRI Transfer Functions In Fourier domain a) TF1 ($18\%$) b)  TF2 ($9\%$)}
 	\centering
 	\includegraphics[width=0.5\textwidth]{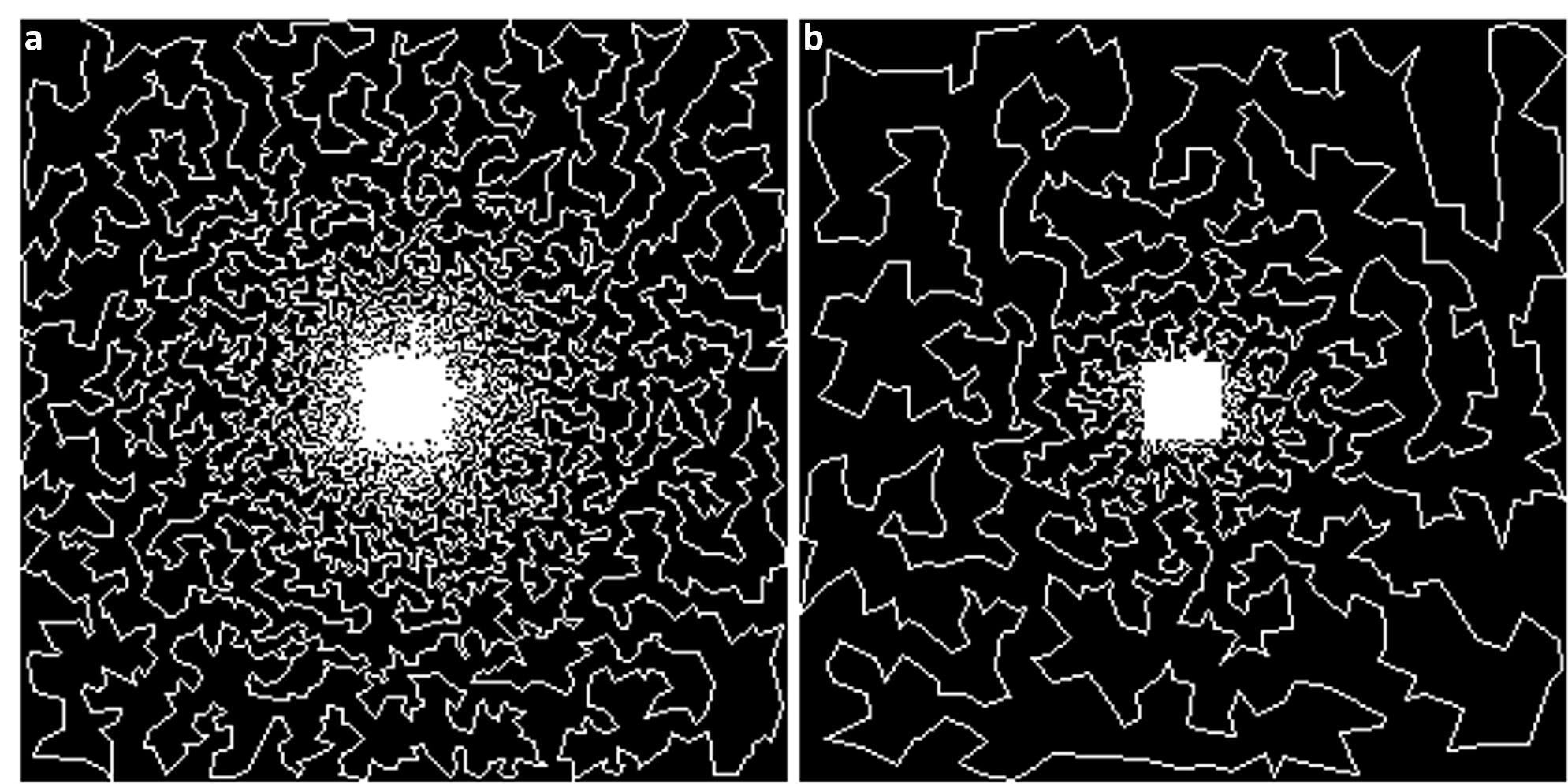}
 	\label{ImageTFall}
 \end{figure}

  Recall that we have two main contributions in this paper:  (i)  a generalization of Hessian-Schatten $p$-norm \cite{hessian} with the resulting form that also generalizes
second order total generalization variation regularization (TGV-2) \cite{tgv};  (ii)  a  novel ADMM based reconstruction algorithm with  improved numerical behaviour
owing to the novel variable splitting scheme.   Implementing the algorithm with novel variable splitting  is enabled by  novel proximal operators derived in
this paper.   The goal of this section is to demonstrate the  role of both of these contributions in improving the quality of reconstruction.  To this end,  we consider
the problem of reconstructing images from quasi-random Fourier samples.  We use the images given in
\cref{Image: All Images} as the models.   To obtain quasi-random Fourier samples,  we use the trajectories generated by solving travelling salesman problem
as proposed by Chauffert  et al  \cite{chauffert2014variable}. 
Sampling trajectories corresponding to two sampling densities were used,  which are
given in \cref{ImageTFall}.  The first one covers $18\%$ of the samples in the Fourier plane whereas the second trajectory covers $9\%$. 
We   restrict $p$ to be in  $\{1,2\}$.

 As done by Chauffert  et al  \cite{chauffert2014variable}, we quantize the sample locations generated from such trajectories 
by the grid corresponding to the DFT of the image.  This helps with the fast implementation of the algorithms.  
Suppose ${\cal T}$ denotes the operation of obtaining the vector of Fourier samples from the quantized Fourier locations,  and let
$\mathbf{m}_f$ be the vector of complex Fourier samples.  Since the noise is Gaussian,  the negative log likelihood of the
measurement model is given by  $\|{\cal T}g - {\mathbf m}_f\|_2^2$.  It can be shown that we have the following relation:
$\|{\cal T}g - {\mathbf m}_f\|_2^2 =   \|h*g - m\|_{2,2}^2$, where $m$ represents inverse DFT of zero-embedded Fourier measurements, ${\mathbf m}_f$,
and $h$  represents inverse DFT of  the image that has ones corresponding to Fourier sampling locations, and zeros everywhere else. 
Note that $h$ and $m$ will typically be complex.  In summary, 
the algebraic form of algorithm developed in Section \ref{sec:costform} matches with the measurement model considered in our experiments.

\subsection{Experiment 1}

In this experiment, the goal is to show the advantage of both the contributions described above.
To this end,  we consider three forms of reconstruction methods that differs from each other
in terms of regularization  and/or the minimization methods:
\begin{itemize}
\item{GHS-1:}  Reconstruction  using the Generalized Hessian Schatten norm with $p=1$  and with proposed ADMM method for minimization
\item{GHS-2:}  Reconstruction  using the Generalized Hessian Schatten norm with $p=2$  and with proposed ADMM method for minimization
\item{GHS-2(G):}  Reconstruction  using the Generalized Hessian Schatten norm with $p=2$   with optimization method  proposed by Guo et al.
\cite{guo2014new}
\end{itemize}
Note that, in GHS-2, the regularization is the same as the TGV-2  \cite{tgv}.  This means,  GHS-2 has  novelty only in terms of optimization used,
whereas GHS-1 has novelty both in terms of the regularization and the optimization method.  GHS-2(G) entirely corresponding the method of
Guo et al \cite{guo2014new} with the part corresponding to the additional wavelet regularization removed.     To evaluate these methods,  we generated test dataset from all given images by using the transfer function TF1.
The Fourier samples were corruption by additive white Gaussian noise with variance 4.
Table 1 compares PSNR score of reconstruction  obtained from all three methods  with 500,  1500, and 10000 iterations.
It is clear from the table that GHS-1  gives the best score with all three cases of number of iterations for most cases of measured images,  and GHS-2   comes next
in PSNR.  The PSNR score of GHS-2(G) is always the lowest.  By considering the fact that GHS-2(G) differs from GHS-2 only by
optimization,  we conclude that the proposed ADMM method is more efficient than the optimization method proposed by Guo et al \cite{guo2014new}.
We further note that,  beyond 1500 iterations, there is no further improvement in the reconstructed image, and  there is always a difference
between the reconstruction obtained by GHS-2 and GHS-2(G).  This confirms that the proposed ADMM method is better conditioned numerically.
Considering the average time required for a single iteration,  all methods take comparable amount of time.
 \begin{table}[h]
\begin{center}
\begin{tabular}{|l|l|l|l|l|}
	\hline
	Image                    & Algorithm      & 500 Iter. & 1500 Iter. & 10000 Iter. \\ \hline
	\multirow{3}{*}{Image 1} & GHS-2(G) & 32.52    & 32.53    & 32.53     \\ \cline{2-5} 
	& GHS-2          & 32.64    & 32.66    & 32.66      \\ \cline{2-5} 
	& GHS-1          & 33.36      & 33.37     & 33.37      \\ \hline
	\multirow{3}{*}{Image 2} & GHS-2(G) & 36.79     & 36.79     & 36.79      \\ \cline{2-5} 
	& GHS-2          & 36.90   & 36.92    & 36.92   \\ \cline{2-5} 
	& GHS-1          & 37.24   & 37.25     & 37.25      \\ \hline
	\multirow{3}{*}{Image 3} & GHS-2(G) & 35.92      & 35.92      & 35.92       \\ \cline{2-5} 
	& GHS-2          & 35.98      & 35.98      & 35.98       \\ \cline{2-5} 
	& GHS-1          & 36.27      & 36.27      & 36.27       \\ \hline
	\multirow{3}{*}{Image 4}                  & GHS-2(G) & 32.88    & 32.88     & 32.88      \\ \cline{2-5} 
	& GHS-2          & 32.89    & 32.92     & 32.92      \\ \cline{2-5} 
	& GHS-1          & 33.19     & 33.21    & 33.21     \\ \hline
	\multirow{3}{*}{Image 5} & GHS-2(G) & 30.11      & 30.22      & 30.22       \\ \cline{2-5} 
	& GHS-2          & 30.14      & 30.24      & 30.24       \\ \cline{2-5} 
	& GHS-1          & 30.00      & 30.27      & 30.27       \\ \hline
	\multirow{3}{*}{Image 6} & GHS-2(G) & 41.77      & 42.12      & 42.12       \\ \cline{2-5} 
	& GHS-2          & 41.87      & 42.15      & 42.15       \\ \cline{2-5} 
	& GHS-1          & 41.61      & 42.13      & 42.13       \\ \hline
\end{tabular}
	\caption{Comparison of PSNR scores yielded by  various method at various number of iteration }
	\end{center}
	\label{tab:exp1}
\end{table}

\subsection{Experiment 2}

\begin{figure*}[h!] 
	\caption{ Comparison of restored images corresponding to image 1,  TF1  and  noise level 1. Images on the right are
	the zoomed-in portions from the images on the left.  Location of the portions is shown as box in the top left image.}
	\centering
	\includegraphics[width=0.8\textwidth]{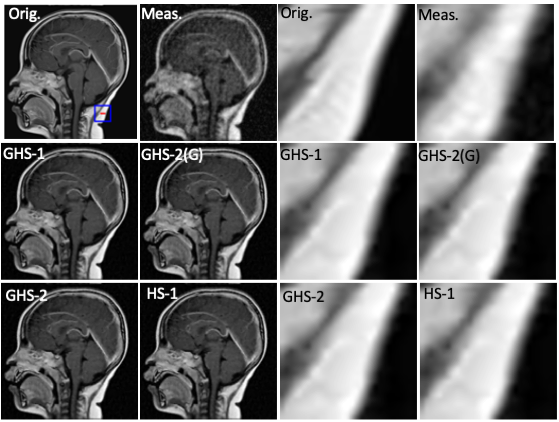}
	\label{Image : Restored Image 1 all}
\end{figure*}

\begin{figure*}[!h]
\caption{Scan lines from the results of figure \ref{Image : Restored Image 1 all}.  The scan line location is shown in figure
\ref{Image : Restored Image 1 all}  as a line segment. }
	\centering
	\includegraphics[width=0.8\textwidth]{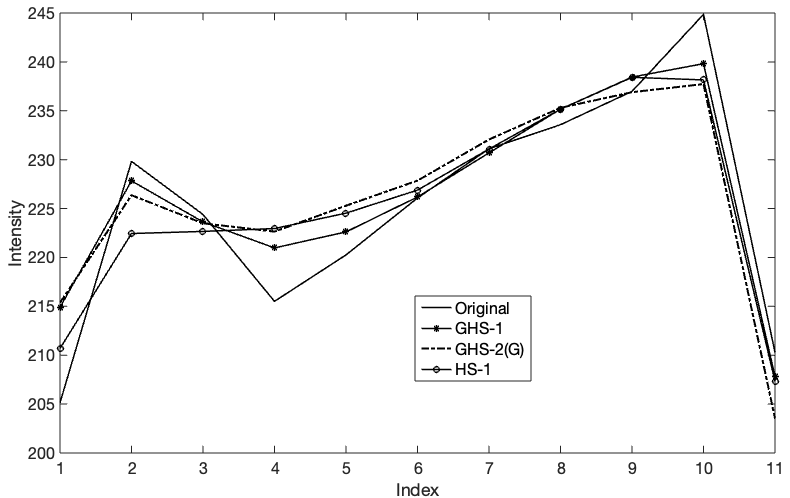}
		\label{Image :Im1_TF1_20_combined}
\end{figure*}

In the second set of experiments, we demonstrate the importance of novel regularization and the novel optimization
under varied input settings.  To this end,  we simulate  measurement data sets using both transfer function 
and add complex Gaussian noise on the Fourier samples with standard deviation values  5 and 7,   which will  be referred to  
as noise levels 1 and 2.   This makes  a total of 24 measurement sets.  We evaluate all three methods listed in the
previous experiment.  We also evaluate with two additional methods:  (i)  HS-1: reconstruction using  Hessian-Schatten 
1-norm regularization with  ADMM based minimization; (ii) HS-2:  reconstruction using  Hessian-Schatten  2-norm
regularization with  ADMM based minimization.  Note that  Hessian-Schatten 2-norm regularization
is also the same as TV-2 regularization.  The results are displayed in Table 2.   From the table,  it is clear that GHS-1 is
the best performing method. We also note that, among the three methods,  GHS-1, GHS-2,  GHS-2(G),  we see the
same pattern of relative performance as in the first experiment.  Further,  GHS-1 and GHS-2 are  better than HS-1 and
HS-2  most cases.  Moreover,  while GHS-2 is consistently  better than   HS-2,  GHS-2(G) is not always better
than  HS-2, although its regularization is the same as that of GHS-2.  As GHS-2  differs from GHS-2(G) only 
by the optimization technique, this again confirm the importance our novel  optimization method. The images restored
from the measurement simulated from image 1 using the transfer function TF1 with noise level 1
 are displayed in    \cref{Image : Restored Image 1 all}.  
We also display a zoomed-in region in each of the restored images. It is clear from the zoomed in images that GHS-1 and 
GHS-2 schemes better recovers the edges without any staircase effect. 
 In    \cref{Image :Im1_TF1_20_combined}, we present scan-lines from this image. As it is difficult to have clarity in the plot
 if several scan-lines are shown, we chose  to show scan-lines  of    reconstructions from three methods only 
 along with the scan-line from
 the original image; we chose the best performing variant from the proposed method,  GHS-1,  and the closest competitors
 from the literature,  GHS-2(G) and HS-1.    The scan-lines  confirms that
GHS-1 follows the  ground truth  better than other methods.

 \begin{table}[]
 \begin{center}
 	\begin{tabular}{|c|c|c| l | l | l | l | l|}
 		\hline
 		\multicolumn{1}{|l|}{Name}     & \multicolumn{1}{l|}{\textbf{TF}} & \multicolumn{1}{l|}{\textbf{Noise Level}} & \textbf{TV2} & \textbf{HS}    & \textbf{GHS-2} & \textbf{GHS-1} & \textbf{GHS-2(G)} \\ \hline
 		\multirow{4}{*}{\textbf{Im 1}} & \multirow{2}{*}{TF1}             & 1                                   & 31.41        & 31.80          & 31.37          & \textbf{31.88} & 31.27                   \\ \cline{3-8} 
 		&                                  & 2                                   & 30.78        & 31.10          & 30.76          & \textbf{31.17} & 30.66                   \\ \cline{2-8} 
 		& \multirow{2}{*}{TF2}             & 1                                   & 26.22        & 26.66          & 26.22          & \textbf{26.68} & 26.10                   \\ \cline{3-8} 
 		&                                  & 2                                   & 26.00        & 26.36          & 26.00          & \textbf{26.44} & 25.88                   \\ \hline
 		\multirow{4}{*}{\textbf{Im 2}} & \multirow{2}{*}{TF1}             & 1                                   & 33.55        & 33.69          & 33.53          & \textbf{33.70} & 33.43                   \\ \cline{3-8} 
 		&                                  & 2                                   & 32.44        & 32.57          & 32.41          & \textbf{32.58} & 32.31                   \\ \cline{2-8} 
 		& \multirow{2}{*}{TF2}             & 1                                   & 29.20        & \textbf{29.37} & 29.15          & 29.30          & 29.08                   \\ \cline{3-8} 
 		&                                  & 2                                   & 28.72        & 28.83          & 28.68          & \textbf{28.83} & 28.60                   \\ \hline
 		\multirow{4}{*}{\textbf{Im 3}} & \multirow{2}{*}{TF1}             & 1                                   & 34.27        & 34.38          & 34.23          & \textbf{34.38} & 34.15                   \\ \cline{3-8} 
 		&                                  & 2                                   & 33.63        & 33.72          & 33.59          & \textbf{33.72} & 33.52                   \\ \cline{2-8} 
 		& \multirow{2}{*}{TF2}             & 1                                   & 28.19        & 28.30          & 28.15          & \textbf{28.34} & 28.11                   \\ \cline{3-8} 
 		&                                  & 2                                   & 28.04        & \textbf{28.12} & 27.96          & 28.10          & 27.92                   \\ \hline
 		\multirow{4}{*}{\textbf{Im 4}} & \multirow{2}{*}{TF1}             & 1                                   & 31.44        & 31.65          & 31.48          & \textbf{31.74} & 31.40                   \\ \cline{3-8} 
 		&                                  & 2                                   & 30.82        & 30.97          & 30.87          & \textbf{31.09} & 30.79                   \\ \cline{2-8} 
 		& \multirow{2}{*}{TF2}             & 1                                   & 26.01        & 26.24          & 26.43          & \textbf{26.47} & 26.35                   \\ \cline{3-8} 
 		&                                  & 2                                   & 25.91        & 26.07          & 26.21          & \textbf{26.22} & 26.13                   \\ \hline
 		\multirow{4}{*}{\textbf{Im 5}} & \multirow{2}{*}{TF1}             & 1                                   & 28.81        & 29.19          & 28.93          & \textbf{29.23} & 28.93                   \\ \cline{3-8} 
 		&                                  & 2                                   & 28.43        & 28.78          & 28.41          & \textbf{28.81} & 28.34                   \\ \cline{2-8} 
 		& \multirow{2}{*}{TF2}             & 1                                   & 22.96        & 23.23          & 23.10          & \textbf{23.25} & 23.08                   \\ \cline{3-8} 
 		&                                  & 2                                   & 22.85        & 23.12          & 22.96          & \textbf{23.16} & 22.92                   \\ \hline
 		\multirow{4}{*}{\textbf{Im 6}} & \multirow{2}{*}{TF1}             & 1                                   & 33.74        & 33.93          & \textbf{34.22} & 34.21          & 34.22                   \\ \cline{3-8} 
 		&                                  & 2                                   & 32.50        & 32.68          & \textbf{32.77} & 32.76          & 32.77                   \\ \cline{2-8} 
 		& \multirow{2}{*}{TF2}             & 1                                   & 28.82        & 28.97          & \textbf{29.31} & 29.30          & 29.29                   \\ \cline{3-8} 
 		&                                  & 2                                   & 28.47        & 28.60          & 28.63          & \textbf{28.63} & 28.62                   \\ \hline
 	\end{tabular}
 \end{center}
\caption{PSNR (dB) of   various reconstruction methods }
\end{table}

\section{Conclusion}

We proposed a new form  of regularization named Generalized Hessian Schatten Norm  (GHSN) regularization. 
GHSN generalizes all existing forms of second-order derivative based regularization.  We also developed a novel
ADMM optimization method  for image reconstruction using GHSN.  We demonstrated the  advantage  of the 
generality in GHSN experimentally. We also demonstrated the effectiveness of the novel optimization method.
In particular,  even when parameters of GHSN is restricted to  such that it becomes the well known form, called
second order total generalized (TGV-2) variation,  our optimization method outperforms the optimization proposed
for TGV-2 in the literature.
 
\section*{Appendix A: Proof of propositions}

\subsection*{Proof of proposition 1}
First we reproduce Eq. \eqref{eq:ghsfp}:
\begin{align}
{\cal GHS}_p(g, \alpha_s, \alpha_f)&= 
\max_{\substack{\|\overline{\mathbf {N}}\|_{\infty,S(q)}\le \alpha_s\\
		\|\overline{\mathbf {N}}*\tilde{\bf d}\|_{\infty,2}\le \alpha_f} }\langle {\overline{\mathbf N}}, {\mathbf H}*g \rangle .
\end{align}
Next, we denote the set $\{ {\mathbf N}|\|\overline{\mathbf N}\|_{\infty,S(q)}\le \alpha_s\}$ as $\mathcal B_S$ and, 
$\{\mathbf z | \|\mathbf z\|_{\infty,2}\le \alpha_f\}$ as $\mathcal B_2$.  This gives
\begin{equation}
{\cal GHS}_p(g, \alpha_s, \alpha_f)= 
\max_{\substack{({\mathbf{N}},\overline{\mathbf{N}}*\tilde{d})\in \mathcal B_{S}\times\mathcal B_2} }\langle {\overline{\mathbf N}}, {\mathbf H}*g \rangle.
\end{equation}
Next, we note that $d_{xx} = d_x*d_x$,  $d_{y} = d_y*d_y$,  and $d_{xy} = d_x*d_y$.  This means that we have 
${\mathbf H}*g = g*{\mathbf d}*{\mathbf d}^t$.     Substituting this gives
\begin{equation}
{\cal GHS}_p(g, \alpha_s, \alpha_f)= 
\max_{\substack{({\mathbf{N}},\overline{\mathbf{N}}*\tilde{d})\in \mathcal B_{S}\times\mathcal B_2} }\langle {\overline{\mathbf N}}, g*{\mathbf d}*{\mathbf d}^t \rangle.
\end{equation}
By the property inner products with convolution,    we can replace the operation $(\cdot)*{\mathbf d}^t$ applied on the second argument of
the inner product by the adjoint operation  $(\cdot)*\tilde{\mathbf{d}}$ applied on the second argument. This gives
\begin{equation}
{\cal GHS}_p(g, \alpha_s, \alpha_f)= 
\max_{\substack{({\mathbf{N}},\overline{\mathbf{N}}*\tilde{d})\in \mathcal B_{S}\times\mathcal B_2} }\langle \overline{\mathbf{N}}*{\mathbf{\tilde{d}}}, {\mathbf{d}}*g \rangle.
\end{equation}
Now, we can replace the maximization by the minimization of the negated function and obtain
\begin{equation}
{\cal GHS}_p(g, \alpha_s, \alpha_f)= -
\min_{\substack{({\mathbf{N}},\overline{\mathbf N}*\tilde{d})\in \mathcal B_{S}\times\mathcal B_2}}-\langle \overline{\mathbf N}*\tilde{\mathbf d}, {\mathbf d}*g \rangle.
\end{equation}

The above minimization problem can be posed as constrained optimization problem as
given below:
\begin{align}
{\cal GHS}_p(g, \alpha_s, \alpha_f) & = -
\min_{\substack{({\mathbf N},\mathbf p)\in \mathcal B_{S}\times\mathcal B_2}} \;\; -\langle {\mathbf p}, {\mathbf d}*g \rangle
\\
\nonumber
& \mbox{subject to} \;\;   \overline{\mathbf N}*\tilde{\mathbf d} = {\mathbf p}.
\end{align}
Since the above problem is a convex optimization problem,    ${\cal GHS}_p(g, \alpha_s, \alpha_f)$
can be determined using duality theory.   To this end,  we construct the Lagrange dual cost function of the minimization problem as given below:
\begin{equation}
\label{eq:ghsdual}
q({\mathbf u}) = \min_{\substack{({\mathbf N},\mathbf p)\in \mathcal B_{S}\times\mathcal B_2}}  \;\;
-\langle {\mathbf p}, {\mathbf d}*g \rangle+ 
\langle {\mathbf u}, {\mathbf p} - \overline{\mathbf N}*\tilde{\mathbf d}  \rangle.
\end{equation}
The above problem can be rearranged  as given below by separating the terms for $\mathbf p$ and $\mathbf N$:
\begin{equation}
\label{eq:ghsdual2}
q({\mathbf u}) = \min_{\substack{({\mathbf N},\mathbf p)\in \mathcal B_{S}\times\mathcal B_2}}  \;\;
-\langle {\mathbf p}, {\mathbf d}*g -{\mathbf u} \rangle
+  
\langle {\mathbf u},  - \overline{\mathbf N}*\tilde{\mathbf d}  \rangle.
\end{equation}
Next, we note that
$\langle {\mathbf u},  - \overline{\mathbf N}*\tilde{\mathbf d}  \rangle 
= \langle {\mathbf u}*{\mathbf d}^t,  - \overline{\mathbf N} \rangle 
=  \langle \overline{{\mathbf u}*{\mathbf d}^t},  - {\mathbf N} \rangle$.  This gives
\begin{equation}
\label{eq:ghsdual3}
q({\mathbf u}) = \min_{\substack{({\mathbf N},\mathbf p)\in \mathcal B_{S}\times\mathcal B_2}}  \;\;
-\langle {\mathbf p}, {\mathbf d}*g -{\mathbf u} \rangle
+
\langle \overline{{\mathbf u}*{\mathbf d}^t},  - {\mathbf N} \rangle.
\end{equation}
By duality theory,  since strong duality holds (\cite{nedic}, Proposition 6.4.2), we have $-{\cal GHS}_p(g, \alpha_s, \alpha_f) = \max_{\mathbf u} q({\mathbf u})$.
Hence,  the cost ${\cal GHS}_p(g, \alpha_s, \alpha_f)$ can be expressed as
\begin{align}
\nonumber &\label{eq:ghsdual3a}
-{\cal GHS}_p(g, \alpha_s, \alpha_f) \\&= \max_{\mathbf u} \;\; \min_{\substack{({\mathbf N},\mathbf p)\in \mathcal B_{S}\times\mathcal B_2}}  \;\;
-\langle {\mathbf p}, {\mathbf d}*g -{\mathbf u} \rangle
+  
\langle \overline{{\mathbf u}*{\mathbf d}^t},  - {\mathbf N} \rangle.
\end{align}
Next,  we rewrite the maximization with respect to ${\mathbf u}$  as the minimization:
\begin{align}
\nonumber \label{eq:ghsdual3b} 
{\cal GHS}_p(g, & \alpha_s, \alpha_f) =\\& \min_{\mathbf u} \;\; \max_{\substack{({\mathbf N},\mathbf p)\in \mathcal B_{S}\times\mathcal B_2}}  \;\;
\langle {\mathbf p}, {\mathbf d}*g -{\mathbf u} \rangle
+  
\langle \overline{{\mathbf u}*{\mathbf d}^t},   {\mathbf N} \rangle.
\end{align}
As  the part of the cost function with respect to ${\mathbf N}$  and ${\mathbf p}$  are separable,
we can rewrite the  respective maximizations independently as  given below:
\begin{align}
& \nonumber \label{eq:ghsdual4}
{\cal GHS}_p(g, \alpha_s, \alpha_f) =\\& \min_{\mathbf u} \;\; \max_{\mathbf N\in \mathcal B_S}  \;\;
\langle \overline{{\mathbf u}*{\mathbf d}^t},   {\mathbf N} \rangle 
+ \max_{\mathbf p\in \mathcal B_2}  
\langle {\mathbf p}, {\mathbf d}*g -{\mathbf u} \rangle.
\end{align}
Now we  apply the definition of conjugate norm.
By noting the fact that
the conjugate (dual) norm for the
norm $\|\cdot\|_{\infty, S(q)} $ is 
$\|\cdot\|_{1, S(p)} $ with $\frac{1}{p}+\frac{1}{q}=1$ \cite{horn2012matrix}, and for $\|\cdot\|_{\infty,2}$ is $\|\cdot\|_{1,2}$.
Substituting this gives
$${\cal GHS}_p(g, \alpha_s, \alpha_f) = \min_{\mathbf u} \;\; \alpha_s \|\overline{{\mathbf u}*{\mathbf d}^t}\|_{1,S(p)}
+ \alpha_f \|{\mathbf d}*g -{\mathbf u}\|_{1,2},$$
which completes the proof.

\subsection*{Proof of proposition 2}
First we note that ${\mathbf A}_f{\mathbf A}_f^t = {\mathbf I}$,  and let ${\mathbf B}_f$ be matrix such that
the augmented matrix ${\mathbf M} = \left[{\mathbf A}_f^t  \;\; \hat{\mathbf B}_f^t\right]^t$ satisfies 
${\mathbf M}^t{\mathbf M} = {\mathbf I}$.  Further, let 
$\zh = {\mathbf M\mathbf z}$, and $\ah = {\mathbf M\mathbf a}$, and  let 
$\zh= \left[
\begin{array}{c}
\zh_1 \\ \zh_2
\end{array}
\right]$ and $\ah= \left[
\begin{array}{c}
\ah_1 \\ \ah_2
\end{array}
\right]$, where 
$\zh_1$  and $\zh_2$  are  sub-vectors  of $\zh$ of size $2\times 1$,  and similarly  
$\ah_1$  and $\ah_2$ sub-vectors  of $\ah$ of size $2\times 1$.
Then the   form of cost function  that is easy to minimize can be obtained 
by substituting 
$\z = {\mathbf M}^t\zh$,  $\a={\mathbf M}^t\ah$,  and $\zh_1 = {\mathbf A}_f\z$  in 
$L_f({\mathbf z}, {\mathbf a}, \alpha_f)$. By doing this, 
we obtain the transformed function as given below:\begin{align}
 \nonumber \hat{L}(\zh_1,\zh_2)=L_f({\mathbf M}^t\zh,  {\mathbf M}^t\ah, \alpha_f) = \\
\frac{\beta}{2}\|\zh_1-\ah_1\|_2^2 + \alpha_f \|\zh_1\|_2 + 
\frac{\beta}{2} \|\zh_2-\ah_2\|_2^2.\end{align}  Let $(\zh_1^*,\zh_2^*)$ denote the minimum of $\hat{L}(\cdot,\cdot)$. 
Clearly,  $\zh_2^* = \ah_2$.   Next,  $\zh_1^*$ is the well-known proximal solution of $l_2$ norm
\cite{parikh2014proximal}, and it is given by $\zh_1^*= \frac{max(\|\ah_1\|_2-t,0)}{\|\ah_1\|_2}\ah_1$, 
where $t=\alpha_f/\beta$. 
From these, the minimum of   $L_f({\mathbf z}, {\mathbf a}, \alpha_f)$, denoted by  $\z^*$ 
can be expressed as $\z^* = {\mathbf M}^t\left[
\begin{array}{c}
\zh_1^* \\ \zh_2^*
\end{array}
\right]  = {\mathbf M}^t\left[
\begin{array}{c}
\zh_1^* \\ \ah_2
\end{array}
\right] $. 
By replacing $\zh_1^*$  by $\ah_1 - (\ah_1-\zh_1^*)$, we get 
$\z^* = {\mathbf M}^t\left[
\begin{array}{c}
\ah_1 \\ \ah_2
\end{array}
\right]- {\mathbf M}^t\left[
\begin{array}{c}
\ah_1-\zh_1 \\ \mathbf 0
\end{array}
\right]= {\mathbf M}^t\left[
\begin{array}{c}
\ah_1 \\ \ah_2
\end{array}
\right]  - {\mathbf A}_f^t(\ah_1-\zh_1^*)$.   Next, from the expression for  $\zh_1^*$, we can deduce
the following on the difference  $(\ah_1-\zh_1^*)$:
\begin{align*}
(\ah_1-\zh_1^*)  &  = \ah_1 -   \frac{max(\|\ah_1\|_2-t,0)}{\|\ah_1\|_2}\ah_1  \\
&  = \frac{1}{\|\ah_1\|}{\ah_1} ( \|\ah_1\|_2 - max( \|\ah_1\|_2-t,0)) \\
& = \frac{1}{\|\ah_1\|}{\ah_1} min( \|\ah_1\|_2, t)
\end{align*}
By using the above relation and by using the fact that 
${\mathbf M}^t\left[
\begin{array}{c}
\ah_1 \\ \ah_2
\end{array}
\right]  = \a$, we get
$$\z^* = \a - \frac{1}{\|\ah_1\|}min( \|\ah_1\|_2, t){\mathbf A}_f^t{\ah_1}.$$
Substituting $\ah_1 = {\mathbf A}_f{\mathbf a}$ in the above expressing gives the final expression.

\subsection*{Proof of proposition 3}

From the definitions of $\K$  and $\L$,  we first note that they can be expressed
as
\begin{equation}
\K = \left[
\begin{array}{cccc}
1 & 0 & 0  & 0\\
0  & 0.5 & 0.5 &  0 \\
0 & 0.5 &  0.5 & 0 \\
0 & 0 & 0 &  1
\end{array}
\right],  \;\;
\L = \left[
\begin{array}{cccc}
0 & 0 & 0  & 0\\
0  & 0.5 & -0.5 &  0 \\
0 & -0.5 &  0.5 & 0 \\
0 & 0 & 0 &  0
\end{array}
\right]
\end{equation}
From the form given above,  we observe that $\K^t = \K$,  $\L = \L^t$,  and $\L\K=\K\L = {\mathbf 0}$.
Hence,   $\frac{1}{2}\|{\mathbf z}-{\mathbf a}\|_2^2$  can be written as  
$\frac{1}{2}\|{\mathbf z}-{\mathbf a}\|_2^2  = \frac{1}{2}\|{\mathbf {Kz}}-{\mathbf {Ka}}\|_2^2  + 
\frac{1}{2}\|{\mathbf {L z}}-{\mathbf {La}}\|_2^2$.  As a result, the minimization problem becomes,  
$$\z^* = 
\underset{\z \in \mathbb{R}^4}{argmin} \;\;
\frac{1}{2}\|{\mathbf {Kz}}-{\mathbf {Ka}}\|_2^2  + 
\frac{1}{2}\|{\mathbf {L z}}-{\mathbf {La}}\|_2^2 +
t \|{\mathbf {K z}}\|_{{S(p)}}.$$
As an additional property,  we also observe that $\K + \L = {\mathbf I}$.   Hence,  we have 
${\cal R}({\mathbf K}) \oplus {\cal R}({\mathbf L}) = 
\mathbb{R}^4$.  This means that 
the minimum of $L_s({\mathbf z}, {\mathbf a}, \alpha_s)$, denoted by
$\z^*$, can be written as $\z^* = \z_k^* + \z_l^*$, where 
\begin{align*}&(\z_k^*, \z_l^*) = 
\underset{\z_k \in {\cal R}({\mathbf K}), \;\;  \z_l \in {\cal R}({\mathbf L})}{argmin} 
\frac{1}{2}\|{\mathbf K}(\z_k+\z_l)-{\mathbf Ka}\|_2^2\\&  + 
\frac{1}{2}\|{\mathbf L }(\z_k+\z_l)-{\mathbf La}\|_2^2 +
t \|{\mathbf K }(\z_k+\z_l)\|_{{S(p)}}.\end{align*}
Next, because of the relation  $\L\K=\K\L = {\mathbf 0}$,  we have 
$\K \z_l = \L \z_k = {\mathbf 0}$.  This means that the minimization subproblem can be separated
as
\begin{align}
\z_k^* &  = 
\underset{\z_k \in {\cal R}({\mathbf K})}{argmin} 
\frac{1}{2}\|{\mathbf K}\z_k-{\mathbf {Ka}}\|_2^2  + 
t \|{\mathbf K }\z_k\|_{{S(p)}}  \\
\z_l^* &  = 
\underset{ \z_l \in {\cal R}({\mathbf L})}{argmin} 
\frac{1}{2}\|{\mathbf L }\z_l-{\mathbf {La}}\|_2^2 
\end{align}	
Next,  we observe that $\K^2={\K}$  and $\L^2 = {\L}$, this means $\K$ and $\L$ are orthogonal projections on range spaces of $\K$ and $\L$ respectively, which implies that
$\K\z_k = \z_k$  and $\L\z_l = \z_l$.   Hence the above minimization problems
can be written as
\begin{align}
\z_k^* &  = 
\underset{\z_k \in {\cal R}({\mathbf K})}{argmin} 
\frac{1}{2}\|\z_k-{\mathbf K\a}\|_2^2  + 
t \|\z_k\|_{{S(p)}}  \\
\z_l^* &  = 
\underset{ \z_l \in {\cal R}({\mathbf L})}{argmin} 
\frac{1}{2}\|\z_l-{\mathbf L\a}\|_2^2 
\end{align}
From the above forms of minimization sub-problems, it  is clear that 
$\z_l^* = \L\a$. We claim that  $\z_k^* = {\cal P}_s(\K\a,t)$  where   ${\cal P}_s(\K\a,t)$  is the proximal
of the Schatten-Norm applied on $\K\a$ with threshold $t$. This is because ${\cal P}_s(\K\a,t)$ is the global unconstrained minimizer of the cost $\frac{1}{2}\|z_k-{\mathbf Ka}\|_2^2  + 
t \|\z_k\|_{{S(p)}}$, also by definition of ${\cal P}_s(\K\a,t)$ \cite{poisson_hessian} (section III.E),  it can  be seen that ${\cal P}_s(\K\a,t)\in \mathcal R(\mathbf K)$. From the above two statements it can be concluded that ${\cal P}_s(\K\a,t)$  is the required minimizer.  Hence the required solution becomes
$\z^* = \L\a +  {\cal P}_s(\K\a,t)$.

\bibliographystyle{IEEEbib}
\bibliography{ghsn_f}

\end{document}